\documentclass[fleqn,usenatbib]{mnras}

\usepackage{newtxtext,newtxmath}
\usepackage{xcolor, soul}
\definecolor{teagreen}{rgb}{0.82, 0.94, 0.75}
\sethlcolor{teagreen}

\usepackage[T1]{fontenc}

\DeclareRobustCommand{\VAN}[3]{#2}
\let\VANthebibliography\thebibliography
\def\thebibliography{\DeclareRobustCommand{\VAN}[3]{##3}\VANthebibliography}

\usepackage{graphicx}	
\usepackage{amsmath}	

\usepackage{stackengine}

\title[WISDOM XX.\ -- 
Strong shear in NGC~524]{WISDOM project XX.\ -- 
Strong shear tearing molecular clouds apart in NGC~524}

\author[A.\ Lu et al.]{Anan Lu,$^{1}$\thanks{E-mail: anan.lu@mail.mcgill.ca}
Daryl Haggard,$^{1}$
Martin Bureau,$^{2}$
Jindra Gensior,$^{3}$
Sarah Jeffreson,$^{4}$
Carmelle Robert,$^{5}$
\newauthor
Thomas G.\ Williams,$^{2}$
Fu-Heng Liang,$^{2}$
Woorak Choi,$^{6}$
Timothy A.\ Davis,$^{7}$
Sara Babic,$^{1}$
Hope Boyce,$^{1}$
\newauthor
Benjamin Cheung,$^{1}$
Laurent Drissen,$^{5}$
Jacob S.\ Elford,$^{7}$
Lijie Liu,$^{8}$
Thomas Martin,$^{5}$
Carter Rhea,$^{9}$
\newauthor
Laurie Rousseau-Nepton$^{10}$ and
Ilaria Ruffa$^{7}$
\\
$^{1}$Trottier Space Institute and Department of Physics, McGill University, 3600 University Street, Montreal, QC H3A~2T8, Canada\\
$^{2}$Sub-department of Astrophysics, Department of Physics, University of Oxford, Denys Wilkinson Building, Keble Road, Oxford OX1~3RH, UK\\
$^{3}$Department of Astrophysics, University of Zurich, Winterthurerstrasse 190, 8057 Z{\"u}rich, Switzerland\\
$^{4}$Harvard-Smithsonian Center for Astrophysics, Harvard University, 60 Garden Street, Cambridge MA, 02138, United States\\
$^{5}$D{\'e}partement de Physique, de G{\'e}nie Physique et d’Optique, Universit{\'e} Laval, Qu{\'e}bec, QC G1V~0A6, Canada\\
$^{6}$Department of Astronomy, Yonsei University, 50 Yonsei-ro, Seodaemun-gu, Seoul 03722, Republic of Korea\\
$^{7}$Cardiff Hub for Astrophysics Research \& Technology, School of Physics \& Astronomy, Cardiff University, Queens Buildings, Cardiff, CF24~3AA,UK\\
$^{8}$DTU-Space, Technical University of Denmark, Elektrovej 327, DK-2800 Kgs.\ Lyngby, Denmark\\
$^{9}$D{\'e}partement de Physique, Universit{\'e} de Montr{\'e}al, Succ.\ Centre-Ville, Montr{\'e}al, Qu{\'e}bec, H3C~3J7, Canada \\
$^{10}$David A.\ Dunlap Department of Astronomy and Astrophysics, University of Toronto, 50 St-George Street, Toronto, M5S~3H4, Canada
\\
}

\date{Accepted XXX. Received YYY; in original form ZZZ}

\pubyear{2024}

\begin{document}
\label{firstpage}
\maketitle

\begin{abstract} 
    Early-type galaxies (ETGs) are known to harbour dense spheroids of stars but scarce star formation (SF). Approximately a quarter of these galaxies have rich molecular gas reservoirs yet do not form stars efficiently. We study here the ETG NGC~524, with strong shear suspected to result in a smooth molecular gas disc and low star-formation efficiency (SFE). We present new spatially-resolved observations of the \textsuperscript{12}CO(2-1)-emitting cold molecular gas from the Atacama Large Millimeter/sub-millimeter Array (ALMA) and of the warm ionised-gas emission lines from SITELLE at the Canada-France-Hawaii Telescope. Although constrained by the resolution of the ALMA observations ($\approx37$~pc), we identify only $52$ GMCs with radii ranging from $30$ to $140$~pc, a low mean molecular gas mass surface density $\langle\Sigma_{\rm gas}\rangle\approx125$~M$_\odot$~pc$^{-2}$ and a high mean virial parameter $\langle\alpha_{\rm obs,vir}\rangle\approx5.3$.
    We measure spatially-resolved molecular gas depletion times ($\tau_{\rm dep}\equiv1/{\rm SFE}$) with a spatial resolution of $\approx100$~pc within a galactocentric distance of $1.5$~kpc. The global depletion time is $\approx2.0$~Gyr but $\tau_{\rm dep}$ increases toward the galaxy centre, with a maximum $\tau_{\rm dep,max}\approx5.2$~Gyr. However, no pure \ion{H}{II} region is identified in NGC~524 using ionised-gas emission-line ratio diagnostics, so the $\tau_{\rm dep}$ inferred are in fact lower limits. Measuring the GMC properties and dynamical states, we conclude that shear is the dominant mechanism shaping the molecular gas properties and regulating SF in NGC~524. This is supported by analogous analyses of the GMCs in a simulated ETG similar to NGC~524.
\end{abstract}



\section{Introduction}
\label{intro}

Results from the past two decades of optical imaging (e.g.\ the Sloan Digital Sky Survey, SDSS; \citealt{2000AJ....120.1579York}) of the local (redshifts $z\lesssim0.1$) Universe have revealed the existence of a bimodality in the distribution of local galaxies: blue star-forming late-type galaxies (LTGs) form a so-called star-forming `main sequence' and red quiescent early-type galaxies (ETGs) form a `red sequence' \citep[e.g.][]{kauffmann_2003MNRAS.341...54K, Cirasuolo2007MNRAS.380..585C}. Despite their name, ETGs are the more evolved systems, having gone through mergers and intense star formation (SF) and now harbouring large spheroids of old stars. The SF rates (SFRs) of ETGs are thus well below those of main-sequence galaxies \citep[e.g.][]{Noeske2007ApJ...660L..43N, Schiminovich2007ApJS..173..315S, Wuyts2011ApJ...742...96W, Salmi2012ApJ...754L..14S}. One well-accepted theory associates insufficient SF with a lack of cold molecular gas. While it is true that many ETGs have had their cold gas depleted and/or heated as a result of merger-triggered starbursts and/or active galactic nucleus (AGN) feedback, recent molecular gas surveys of nearby galaxies raise another question: why is SF quenched in gas-rich ETGs? In the local Universe, $\approx23\%$ of ETGs harbour substantial molecular gas reservoirs \citep{Young2011MNRAS.414..940Y, Davis2019MNRAS.486.1404D}. This gas is typically located in centrally-concentrated discs that are dynamically cold \citep{Alatalo2013MNRAS.432.1796A, Davis2013MNRAS.429..534D, Ruffa2019MNRAS.484.4239R, 2019MNRAS.489.3739Ruffa}. To preserve the quenched SF observed in these ETGs, the physical conditions of the cold molecular gas reservoirs must be significantly different from those of LTGs. 

The cold molecular gas in the interstellar medium (ISM) is closely linked to ongoing SF \citep[e.g.][]{Leroy2008AJ....136.2782L} and thus the evolution of galaxies. Molecular gas in present-day galaxy discs is formed from the dense atomic gas, compressed by external forces (e.g.\ spiral density waves) and/or its own gravity via local instabilities. Under the correct conditions (e.g.\ shielded from hard radiation), cold molecular gas fragments into giant molecular clouds (GMCs) that are the birthplaces of star clusters.
The properties of GMCs are governed by mechanisms that clump gas together as well as dissipative forces that tend to break clouds apart (e.g.\ shear and turbulence). 
Therefore, quantitative analyses of GMC properties and their governing mechanisms should disentangle what separates ETGs from LTGs in terms of (quenched) SF \citep[see e.g.][]{chevance2022life}.

The first systematic study of GMCs in an ETG was carried out by \citet{Utomo2015ApJ...803...16U} using Combined Array for Research in Milimeter-wave Astronomy observations of NGC~4526.
These authors showed that the basic properties (e.g.\ size, linewidth or velocity dispersion and luminosity) of the GMCs of this ETG 
deviate from those of Milky Way (MW) GMCs. NGC~4526 GMCs have a much larger linewidth and luminosity at a given radius and no size -- linewidth correlation. Another ETG, NGC~4429, was studied by \citet{Liu2021MNRAS.505.4048L} and has smaller, denser and more elongated GMCs than the MW. The properties of the NGC~4429 GMCs are shaped by galactic shear.
However, a similar ETG, NGC~1387, has GMCs that resemble those of the MW disc \citep{Liang1387inprep}. \citet{williams2023wisdom} recently performed a beam-by-beam analysis of the molecular gas of $7$ ETGs as part of the mm-Wave Interferometric Survey of Dark Object Masses (WISDOM; \citealt{2017MNRAS.468.4663Onishi}). Molecular gas in these ETGs has higher velocity dispersions, virial parameters and internal turbulent pressures than those of nearby star-forming galaxies. As these works have established, while there are variations, the properties of ETG GMCs are generally different from those of GMCs in the MW and nearby star-forming galaxies.

ETGs are characterised by large stellar spheroids. This type of centrally-concentrated stellar distribution creates a deep gravitational potential well that leads to a sharply rising circular velocity curve in the inner regions \citep[e.g.][]{Yoon2021ApJ...922..249Y}{}{}, and in turn to strong shear. This is known to impact GMC shapes and properties \citep[e.g.][]{Liu2021MNRAS.505.4048L} and to be associated with a low star-formation efficiency (SFE; \citealt{Davis2014MNRAS.444.3427D}). Large stellar mass volume density gradients, which are the origin of strong shear, are also associated with smoother molecular gas discs \citep{Davis2022MNRAS.512.1522D}.
The influence of strong external forces, such as shear and turbulent pressure, on ETG GMCs has also been confirmed by simulations \citep[e.g.][]{2009ApJ...707..250Martig, gensior2020heart,gensior2023wisdom}.

Here we present a case study of the galaxy NGC~524, an ETG with a very centrally-concentrated stellar mass distribution. We use new spatially-resolved observations of the \textsuperscript{12}CO(2-1)-emitting cold molecular gas (spatial resolution $\approx30$~pc) from the Atacama Large Millimeter/sub-millimeter Array (ALMA) and of the warm ionised-gas emission lines (resolution $\approx100$~pc)  from the imaging Fourier transform spectrograph SITELLE at the Canada-France-Hawaii Telescope (CFHT) to study the GMC properties and SFEs. We present our observational data in Section~\ref{data} while we identify GMCs and discuss their properties and SFEs in Section~\ref{results}. We make a comparison to a simulated ETG in Section~\ref{dis:sim} and discuss the unique role of strong shear to quench SF in Section~\ref{discussion}. We summarise and conclude in Section~\ref{conclusion}.


\section{Target and data}
\label{data}


\subsection{NGC~524}
\label{data:target}

\begin{figure}
  \centering
  \includegraphics[width=0.41\textwidth]{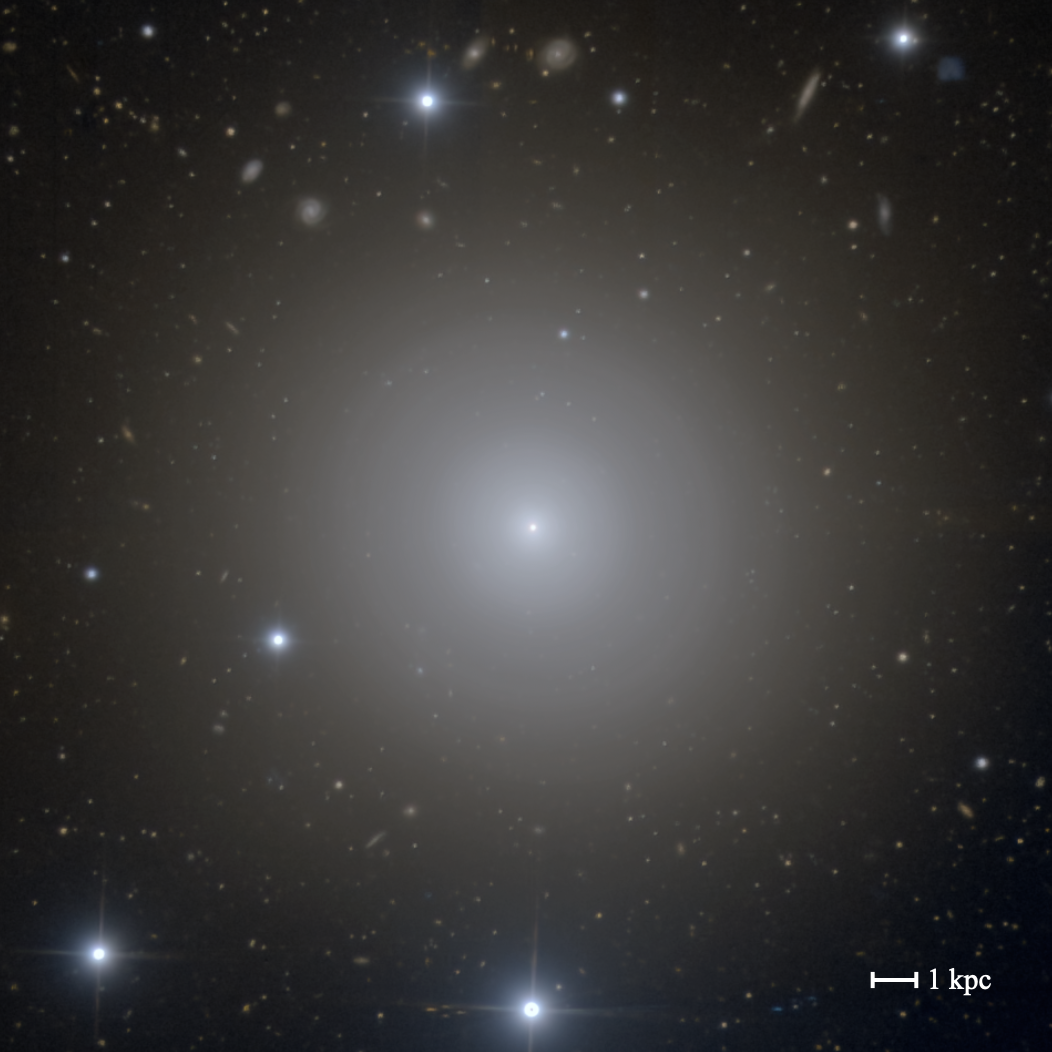}
  \caption{\label{fig:opt} Deep optical image of NGC~524 extracted from our CFHT SITELLE observations. A scale bar is shown in the bottom-right corner.}
\end{figure}

NGC~524 is a nearly face-on SA0 ETG with a cored stellar light profile \citep{Faber1997AJ....114.1771F}. It has an $I$-band effective radius $R_{\rm e}=51\farcs0$ ($\approx5.7$~kpc) and a stellar velocity dispersion within $1$~$R_{\rm e}$ $\sigma_{\rm e}=220$~km~s$^{-1}$ \citep{Cappellari2006MNRAS.366.1126C, Cappellari2013MNRAS.432.1709C}. It is a fast rotator, with a systemic velocity of $2350$~km~s$^{-1}$ and a specific angular momentum within $1~R_{\rm e}$ $\lambda_{R_{\rm e}}=0.28$ \citep{Emsellem2007MNRAS.379..401E}. NGC~524 harbours a dense stellar bulge (shown in Figure~\ref{fig:opt}) and a regular central dust disc with flocculent spiral arms, visible in absorption in {\it Hubble Space Telescope} ({\it HST}) images \citep{Silchenko2000AJ....120..741S, crocker2011MNRAS.410.1197C}. Throughout this paper we adopt a distance to NGC~524 $D=23.3\pm2.3$~Mpc, as used in previous studies \citep[e.g.][]{Smith2019MNRAS.485.4359S}, derived using surface brightness fluctuations \citep{Tonry2001ApJ...546..681T} with the Cepheid zero-point of \citet{Freedman2001ApJ...553...47F}. At this distance, $1\farcs0$ corresponds to $\approx113$~pc. NGC~524 has a total stellar mass of $2.51\times10^{11}$~M$_\odot$, SFR of $0.27$~M$_\odot$~yr$^{-1}$ and molecular gas mass of $8.91\times10^7$~M$_\odot$ \citep{Davis2022MNRAS.512.1522D}. This unique combination of high stellar mass, relatively low molecular gas mass and low SFR makes NGC~524 an ideal system in which to study the influence of the gravitational potential on GMCs and SF.

NGC~524 exhibits nuclear activity and is revealed as a compact radio source at $5$~GHz by the Very Large Array \citep{Nyland2016MNRAS.458.2221N} and Very Long Baseline Interferometry \citep{Filho2004A&A...418..429F}, with a $1.4$~mJy core \citep{1991AJ....101..148Wroble}. However, NGC~524 is under-luminous and does not show any nuclear activity at X-ray wavelengths \citep{2006ApJ...653L...9Dxray}. Based on stellar kinematics obtained with the adaptive optics-assisted Gemini-North telescope, \citet{Krajnovi2009MNRAS.399.1839K} inferred a supermassive black hole (SMBH) mass $M_{\rm BH}=8.3^{+2.7}_{-1.3}\times10^8$~M$_\odot$. They also inferred a stellar mass-to-light ratio in the $I$-band $M/L_{\rm I}=5.8\pm0.4$~M$_\odot$/L$_{\odot,I}$, having assumed an inclination of $20^\circ$ from \citet{Cappellari2006MNRAS.366.1126C}. \citet{Smith2019MNRAS.485.4359S} subsequently used ALMA \textsuperscript{12}CO(2-1) observations to dynamically infer a SMBH mass of $4.0^{+3.5}_{-2.0}\times10^8$~M$_\odot$, roughly consistent with the earlier measurement. 
 

\subsection{Molecular gas observations}
\label{data:CO}

NGC 524’s molecular gas was first observed as a follow-up to the SAURON \citep{2002MNRAS.329..513DSAURONI} and ATLAS$^{\rm 3D}$ \citep{2011MNRAS.413..813C_ATLAS3DI} projects. Observations of both the \textsuperscript{12}CO(2-1) and \textsuperscript{12}CO(1-0) lines with the Institut de Radioastronomie Millim{\'e}trique (IRAM) 30-m telescope revealed a double-horned profile typical of a rotating disc and a total molecular hydrogen mass of $(9\pm1)\times10^7$~M$_\odot$ \citep{2011MNRAS.414..940Young}. Spatially-resolved observations were also obtained using the IRAM Plateau de Bure Interferometer, with a resolution of $2\farcs8\times2\farcs6$ ($\approx320\times290$~pc$^2$), showing that NGC~524 has a fast-rotating molecular gas disc with a total molecular hydrogen mass of $\approx6.7\times10^7$~M$_\odot$ and a maximum radius of $1.1$~kpc \citep{2011MNRAS.410.1197Crocker}.

The NGC~524 molecular gas data used here were obtained with  ALMA using both the 12-m array and 7-m Atacama Compact Array (ACA, also known as the Morita array; \citealt{2009PASJ...61....1Iguchi}). Data were taken as part of the WISDOM project observing programmes 2015.1.00466.S (PI: Onishi), 2016.2.00053.S (PI: Liu) and 2017.1.00391.S (PI: North). These data were first used to measure the supermassive black hole mass at the centre of NGC~524 by \citet{Smith2019MNRAS.485.4359S}. The 12-m data span baselines from $15$~m to $1.3$~km, providing the high spatial resolution required for our project, and were taken in four tracks on 26 March 2016, 17 July 2016, 2 May 2017 and 16 September 2018. The ACA observations span shorter baselines from $9$ to $48$~m, providing the $uv$-plane coverage necessary to map more extended gas structures, 
and were taken in a single track on 25 June 2017. The total on-source time achieved was $2.2$~hr with the 12-m array and $0.3$~hr with the ACA. 

For both arrays, a spectral window was positioned to observe the redshifted $J=2-1$ transition of \textsuperscript{12}CO at a velocity resolution of $\approx1$~km~s$^{-1}$ over a bandwidth of $\approx2500$~km~s$^{-1}$. Three additional spectral windows were positioned to observe the continuum emission, each with a bandwidth of $2$~GHz and a lower velocity resolution.

\begin{figure*}
  \centering
  \includegraphics[width=0.8\textwidth]{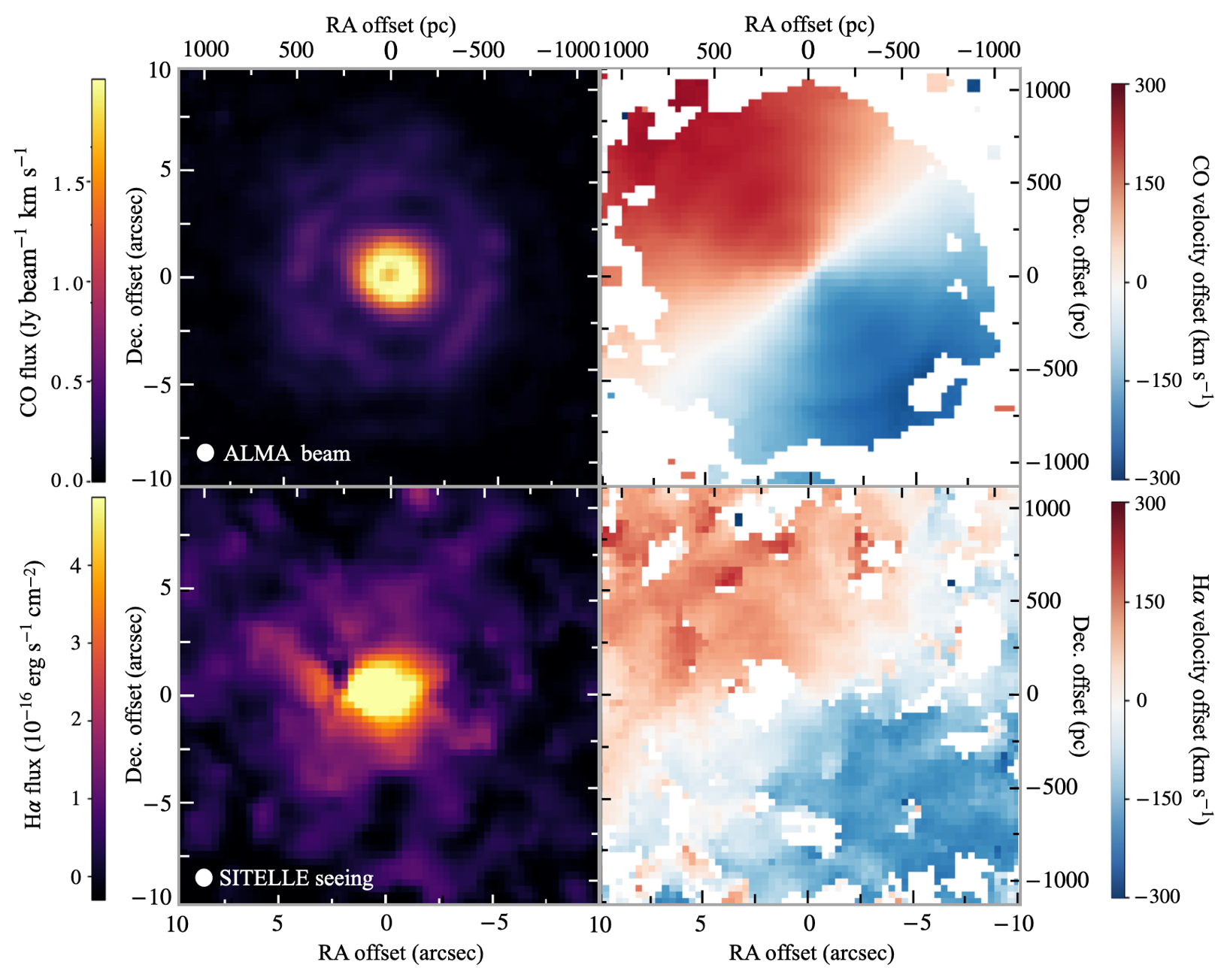}
  \caption{\label{fig:data} ALMA and SITELLE observations of NGC~524, illustrating that this galaxy has a fast-rotating molecular gas disc with two rings but diffuse ionised gas. {\it Top:} \textsuperscript{12}CO(2-1) surface brightness (left) and intensity-weighted mean line-of-sight velocity offset (right) map at the spatial resolution of the SITELLE data. {\it Bottom:} H$\alpha$ surface brightness (left) and intensity-weighted mean line-of-sight velocity offset (right) map, with a seeing of $1\farcs1$ ($\approx125$~pc). The surface brightness maps of other emission lines are shown in Appendix~\ref{app_Halpha}. The velocity offsets assume a systemic velocity of $2350$~km~s$^{-1}$ (see Section~\ref{data:target}). The synthesised beam/seeing is shown in the bottom-left corner of the left-hand maps as a solid white circle ($1\farcs1\times1\farcs1$ or $\approx125\times125$~pc$^2$).}
\end{figure*}

\begin{figure}
  \centering
  \includegraphics[width=0.43\textwidth]{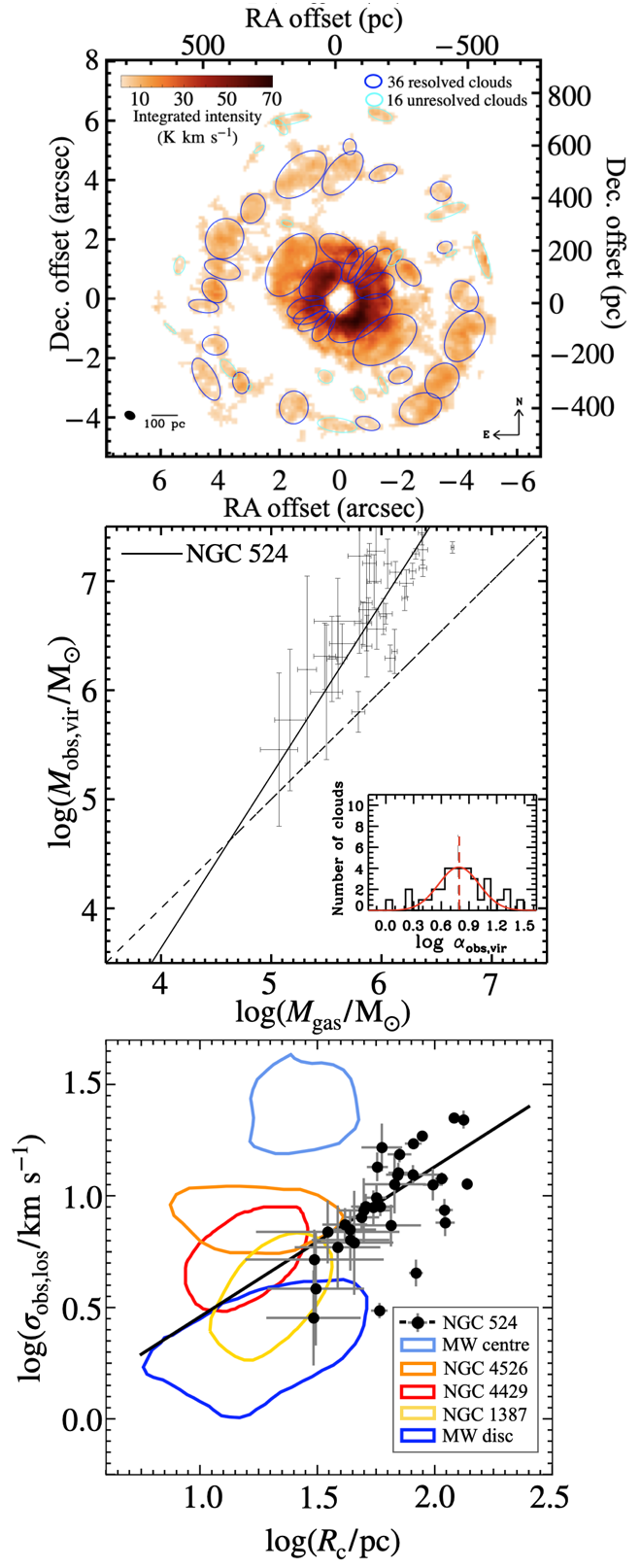}
  \caption{\label{fig:gmc} GMCs of NGC~524, with high virial parameters and a size -- linewidth relation slightly steeper than that of MW disc clouds. {\it Top:} Identified GMCs overlaid on the \textsuperscript{12}CO(2-1) moment-0 map. The synthesised beam is shown in the bottom-left corner as a solid black ellipse ($0\farcs38\times0\farcs29$ or $\approx43\times32$~pc$^2$) along with a scale bar. {\it Middle:} Correlation between virial mass and molecular gas mass for the identified GMCs. The black solid line shows the best-ﬁtting power-law relation, while the black dashed diagonal line indicates the $1:1$ relation. The inset shows the distribution of $\log(\alpha_{\rm obs,vir})$ (black histogram) with a log-normal ﬁt overlaid (red solid line). The red dashed line indicates the mean of the log-normal ﬁt. {\it Bottom:} Size -– linewidth relation of the identidied GMCs, using the observed velocity dispersion $\sigma_{\rm obs,los}$. The black solid line shows the best-ﬁtting power-law relation, while the coloured contours encompass $68\%$ of the data point distributions of the MW (centre and disc) clouds and ETG clouds (NGC~4526, NGC~4429 and NGC~1387). 
  }
\end{figure}

The data were calibrated and combined using the {\tt Common Astronomy Software Applications} ({\tt CASA}; \citealt{McMullin2007ASPC..376..127M}) pipeline. To remove the continuum emission from the line spectral window, 
a linear fit was made to the line-free channels at both ends of that window as well as the three pure continuum spectral windows and was subtracted from the $uv$ plane using the {\tt CASA} task {\tt uvcontsub}. The resulting line data were imaged into two RA-Dec.-velocity cubes with different synthesised beam sizes and channel widths. A first cube was created for the GMC analyses of this paper, with $2$~km~s$^{-1}$ channels and Briggs' weighting with a robust parameter of $0.5$, resulting in a synthesised beam of $0\farcs38\times0\farcs29$ ($\approx43\times32$~pc$^2$) full-width at half-maximum (FWHM) and a sensitivity of $1.1$~mJy~beam$^{-1}$ per channel. A second data cube was created for the SFE studies, with $10$~km~s$^{-1}$ channels, Briggs' weighting with a robust parameter of $1.5$ and a $uv$-taper of $0\farcs9$, thus achieving a synthesised beam of $0\farcs99\times0\farcs98$ ($\approx110\times110$~pc$^2$), roughly matching the seeing of the H$\alpha$ observations (see Section~\ref{data:Halpha}). We then convolved this data cube spatially with a narrow and slightly elongated two-dimensional Gaussian, to achieve a perfectly circular synthesised beam of $1\farcs1\times1\farcs1$ ($\approx125\times125$~pc$^2$) and a sensitivity of $0.3$~mJy~beam$^{-1}$ per channel, that perfectly matches the seeing of the H$\alpha$ observations. The moment-0 (total intensity) and moment-1 (intensity-weighted mean line-of-sight velocity) map of this data cube are shown in Figure~\ref{fig:data}, while the moment-0 map of the $2$~km~s$^{-1}$ channel data cube is illustrated in the top panel of Figure~\ref{fig:gmc}.

The \textsuperscript{12}CO(2-1) flux ($F_{\rm CO(2-1)}$) within each spaxel is obtained from the $1\farcs1\times1\farcs1$ resolution moment-0 map by dividing the surface brightness in each spaxel (in units of Jy~beam$^{-1}$~km~s$^{-1}$) by the synthesised beam area in spaxels. This flux is then converted into a luminosity according to
\begin{equation}
  \frac{L_{\rm CO(2-1)}}{\rm K~km~s^{-1}~pc^2}=\left(\frac{3.25\times10^7}{{(1+z)}^3}\right)\,\left(\frac{F_{\rm CO(2-1)}}{\rm Jy~km~s^{-1}}\right)\,{\left(\frac{\nu_{\rm obs}}{\rm GHz}\right)}^{-2}\,\left(\frac{D}{\rm Mpc}\right)^2
\end{equation}
\citep[e.g.][]{decarli2016alma}, where $z$ is the galaxy redshift and $\nu_{\rm obs}$ is the observed frequency (i.e.\ the redshifted frequency of the \textsuperscript{12}CO(2-1) line). 
The luminosity-based molecular gas mass within each spaxel is then calculated using
\begin{equation}
  \frac{M_{\rm mol}}{{\rm M}_\odot}=4.4\,\left(\frac{L_{\rm CO(1-0)}}{\rm K~km~s^{-1}~pc^2}\right)\,\left(\frac{X_{\rm CO(1-0)}}{\rm 2\times10^{20}\,cm^{-2}~(K~km~s^{-1})^{-1}}\right)\,\,\,,
\end{equation}
where $L_{\rm CO(1-0)}$ is the \textsuperscript{12}CO(1-0) luminosity and $X_{\rm CO(1-0)}$ is the \textsuperscript{12}CO(1-0)-to-molecules conversion factor. We adopt a \textsuperscript{12}CO(2-1)/\textsuperscript{12}CO(1-0) ratio of $0.8$ (in brightness temperature units), typical of spiral galaxies \citep[e.g.][]{lamperti2020co}, and $X_{\rm CO(1-0)}=2.3\times10^{20}$~cm$^{-2}$~(K~km~s$^{-1}$)$^{-1}$, commonly used in extragalactic studies \citep[e.g.][]{hughes2013probability, utomo2015giant, sun2018cloud}, although the latter can depend on the metallicity and environment of a molecular cloud (e.g.\ radiation field; \citealt{bolatto2013co}). The mass contribution of helium and other heavy elements is included in this coefficient \citep{strong1988radial, bolatto2013co}. The complete conversion thus becomes
\begin{equation}
  \frac{M_{\rm mol}}{{\rm M}_\odot}=6.325\left(\frac{3.25\times10^7}{{(1+z)}^3}\right)\,\left(\frac{F_{\rm CO(2-1)}}{\rm Jy~km~s^{-1}}\right)\,{\left(\frac{\nu_{\rm obs}}{\rm GHz}\right)}^{-2}\,\left(\frac{D}{\rm Mpc}\right)^2\,\,\,.
\end{equation}

The molecular gas mass surface density within one spaxel ($\Sigma_{\rm mol}$) is then calculated as $M_{\rm mol}$ divided by the spaxel area. The molecular gas mass surface density within a region tightly encompassing the molecular gas disc (a $20\arcsec\times20\arcsec$ or $\approx2.26\times2.26$~kpc$^2$ box centred at the galaxy centre; see Figure~\ref{fig:data}) is $\Sigma_{\rm mol}=11.5\pm1.3$~M$_\odot$~pc$^{-2}$, calculated as the sum of $M_{\rm mol}$ divided by the area of the region. Within a $1$~kpc galactocentric distance circular aperture, we measure $\Sigma_{\rm mol}=28.9\pm1.3$~M$_\odot$~pc$^{-2}$, 
consistent with the molecular hydrogen mass surface density within that same aperture reported by \citet{Davis2022MNRAS.512.1522D}.

One outstanding feature of the molecular gas disc of NGC~524 is the dual ring structure at galactocentric distances of $\approx150$ and $\approx500$~pc (see Figure~\ref{fig:data}). These rings are molecular gas overdensities that impact the GMC distribution, properties and depletion times (see Sections~\ref{GMC} and \ref{SFE}).


\subsection{Ionised-gas observations}
\label{data:Halpha}

NGC~524 was observed at CFHT with SITELLE \citep{drissen2019sitelle}, an optical imaging Fourier transform spectrograph equipped with two E2V detectors each with $2048\times2064$ pixels. The SITELLE field of view is $11\arcmin\times11\arcmin$, resulting in a mean spaxel size on the sky of $0\farcs31\times0\farcs31$ ($\approx28\times28$~pc$^2$). Two data cubes were obtained: one centred on the emission lines of [\ion{N}{ii}]$\lambda6548$, H$\alpha$, [\ion{N}{ii}]$\lambda6583$, [\ion{S}{ii}]$\lambda6716$ and [\ion{S}{ii}]$\lambda6731$ with the SN3 filter ($6480$ -- $6860$~\AA) at a mean spectral resolution $R\approx2000$ (2022B semester, programme number 22Bc09); the other centred on the emission lines of H$\beta$, [\ion{O}{iii}]$\lambda4959$ and [\ion{O}{iii}]$\lambda5007$ with the SN2 filter ($4840$ -- $5120$~\AA) at a mean spectral resolution $R\approx900$ (2020B semester, programme number 20Bc25). The observations were centred at ${\rm RA~(J2000)}=10^{\rm h}14^{\rm m}15\fs05$ and ${\rm Dec.~(J2000)}=3^\circ27\arcmin57\farcs90$. 

The data reduction was performed with the {\tt ORBS} software developed for SITELLE \citep{martin2015orbs, martin2021data}. The seeing was measured to be $1\farcs1$ ($\approx125$~pc) from the FWHM of Gaussian fits to foreground stars from the Gaia catalogue \citep{lindegren2018gaia}. The SN3 data were further calibrated in wavelength based on velocity measurements of the OH sky line, allowing to use this cube for line-of-sight velocity measurements with an absolute precision of a few km~s$^{-1}$ \citep{martin2016optimal}. Sky subtraction was performed using a median sky spectrum extracted from a $200\times200$ spaxels region located far away from the galaxy.

NGC~524 is an ETG with a high-surface brightness core of old stars, so to measure faint emission-line fluxes, additional care must be taken to accurately remove the strong stellar continuum. At each spaxel and for each of the SN3 and SN2 filters, we integrate spectra within a nearby region allowing for the best signal-to-noise ratio ($S/N$; see Appendix~\ref{app_Halpha}), fit the stellar continuum of this region using penalised pixel fitting ({\tt pPXF}; \citealt{Cappellari2022}) and then subtract this stellar continuum from the original spectrum. This process is critical for the accurate fitting of emission-line fluxes in the SN3 data cube, as illustrated in Appendix~\ref{app_Halpha}. For the SN2 data cube, larger binning is required to achieve sufficiently high $S/N$.

After subtracting the stellar continuum from the spectrum of each spaxel in each filter, the emission lines were fitted using the extraction software {\tt ORCS} \citep{martin2015orbs}. For each emission line, {\tt ORCS} outputs parameters (and their uncertainties) including the integrated flux, peak flux, intensity-weighted mean line-of-sight velocity and FWHM, and continuum level. From these, maps of flux, intensity-weighted mean line-of-sight velocity and intensity-weighted line-of-sight velocity dispersion are generated. A detection threshold is then applied to each spaxel based on the $3\sigma$ noise level of the summed H$\alpha$ and [\ion{N}{ii}] flux. The H$\alpha$ surface brightness map is shown in Figure~\ref{fig:data}, while the \ion{N}{II} and \ion{S}{II} surface brightness maps are shown in Appendix~\ref{app_Halpha}. The ionised gas of NGC~524 is diffuse, without the ring structures present in the molecular gas map.

Due to the low $S/N$ of the SN2 data cube, we cannot use the H$\beta$ fluxes of individual spaxels for extinction-correction purposes. Thus, we identify H$\alpha$ emission peaks and associated regions (analogous to \ion{H}{II} regions, but mixed with the diffused gas), spatially bin the SN2 and SN3 spectra within these regions and fit for H$\beta$ and H$\alpha$ (see Appendix~\ref{app_Halpha}). Assuming a constant ratio of H$\alpha$ ($F_{\rm H\alpha,obs}$) to H$\beta$ ($F_{\rm H\beta,obs}$) flux within each region, a map of $F_{\rm H\alpha,obs}/F_{\rm H\beta,obs}$ is generated, which is used to calculate the colour excess of H$\alpha$ over H$\beta$:
\begin{equation}
  E({\rm H}\beta-{\rm H}\alpha)\equiv2.5\log\left(\frac{(F_{\rm H\alpha,obs}/F_{\rm H\beta,obs})}{({\rm H\alpha}/{\rm H\beta})_{\rm intrinsic}}\right)\,\,\,,
\end{equation}
where we assume $({\rm H\alpha}/{\rm H\beta})_{\rm intrinsic}=2.86$, as expected for case~B recombination at a temperature of $10^4$~K \citep{osterbrock2006astrophysics}. The H$\alpha$ extinction is then calculated as
\begin{equation}
  A_{\rm H\alpha}=\left(\frac{E({\rm H\beta}-{\rm H\alpha})}{k(\lambda_{\rm H\beta})-k(\lambda_{\rm H\alpha})}\right)\,k(\lambda_{\rm H\alpha})
\end{equation}
following \citet{nelson2016spatially}, where $k(\lambda)$ is the reddening curve of \citet{fitzpatrick1986average} and $k(\lambda_{\rm H\alpha})$ and $k(\lambda_{\rm H\beta})$ are evaluated at the wavelengths of H$\alpha$ and H$\beta$, respectively. We show $A_{\rm H\alpha}$ as a function of galactocentric distance in Appendix~\ref{app_Halpha}. There are a few regions with negative $A_{\rm H\alpha}$, potentially due to different ionisation conditions in the diffused gas and uncertainties of the $F_{\rm H\beta,obs}$ measurements. For these regions, we use the upper limit of $A_{\rm H\alpha}$ that is positive. Finally, the extinction-corrected H$\alpha$ flux ($F_{\rm H\alpha}$) is calculated as
\begin{equation}
  F_{\rm H\alpha}=F_{\rm H\alpha,obs}\,e^{A_{\rm H\alpha}/1.086}\,\,\,.
\end{equation}
We show the extinction-corrected H$\alpha$ surface brightness map of NGC~524 in Appendix~\ref{app_Halpha} and as contours in the top panel of Figure~\ref{fig:sfe}.

This extinction-corrected H$\alpha$ flux is converted to a SFR using the relation of \citet{kennicutt2012star}:
\begin{equation}
  \log({\rm SFR}\,/\,{\rm M}_\odot\,{\rm yr}^{-1})=\log(L_{\rm H\alpha}\,/\,{\rm erg}\,{\rm s}^{-1})-41.27\,\,\,,
\end{equation}
where $L_{\rm H\alpha}=F_{\rm H\alpha}(4\pi D^2)$ is the extinction-corrected H$\alpha$ luminosity. We note that when probing spatial scales smaller than $\approx500$~pc, this conversion relation can break down, as seen in examples of \citet{kennicutt2012star}. Local SFRs depend on the environment and age of the stellar population, that we do not consider here. Nevertheless, this conversion holds true for the radial profile of SFE (see Section~\ref{SFE}), which is calculated within apertures sufficiently large for confident H$\alpha$-to-SFR conversions.

The surface density of SFR ($\Sigma_{\rm SFR}$) within one spaxel is then calculated as the SFR within that spaxel divided by the spaxel area. The SFR surface density within the $20\arcsec\times20\arcsec$ ($\approx2.26\times2.26$~kpc$^2$) region tightly encompassing the molecular gas disc (see Figure~\ref{fig:data}) is $\Sigma_{\rm SFR}=(5.6\pm1.8)\times10^{-3}$~${\rm M_\odot~yr^{-1}~kpc^{-2}}$, calculated as the sum of the SFRs divided by the area of the region (see Section~\ref{data:CO}). 
This $\Sigma_{\rm SFR}$ is slightly larger than that reported by \citet{Davis2014MNRAS.444.3427D} within an area of $3.76\pm0.97$~kpc$^2$ ($\Sigma_{\rm SFR}=(4.4\pm1.4)\times10^{-3}$~${\rm M_\odot~yr^{-1}~kpc^{-2}}$), measured from integrated {\it Wide-field Infrared Survey Explorer} $22$~$\mu$m and {\it Galaxy Evolution Explorer} far-ultraviolet (FUV) flux densities. 

We note that the SFRs calculated here are really upper limits, as large fractions of the H$\alpha$ fluxes arise from diffuse emission that can not be characterised as arising from star-forming regions. This is confirmed by the ionised-gas emission-line ratios of [\ion{O}{iii}]/H$\beta$ and [\ion{N}{ii}]/H$\alpha$ (see Appendix~\ref{app_Halpha}), that are not typical of star-forming regions. 


\section{Results: smooth gas disc and suppressed SF}
\label{results}


\subsection{Molecular gas properties}
\label{GMC}

NGC~524 is characterised by a high stellar mass surface density ($5.62\times10^9$~${\rm M_\odot~kpc^{-2}}$, \citealt{Davis2022MNRAS.512.1522D}) and a low molecular gas mass surface density ($28.9$~${\rm M_\odot~pc^{-2}}$) within a galactocentric distance of $1$~kpc. The molecular gas disc is also relatively smooth, as indicated by the smoothness index $S$, that quantifies the smoothness of a two-dimensional distribution on a given spatial scale \citep{Conselice2003ApJS..147....1C}, and the Gini coefficient $G$, that quantifies the (in)equality of a distribution \citep{Abraham1996MNRAS.279L..47A}. \citet{Davis2022MNRAS.512.1522D} reports $S=0.19$ and $G=0.43$ for the \textsuperscript{12}CO(2-1) distribution of NGC~524. Both parameters are at the low end of the ETG sample distribution, indicating that NGC~524 has a smooth and uniform molecular gas distribution. Here we nevertheless attempt to identify GMCs in NGC~524 and analyse their properties.

\subsubsection{GMC identification}
\label{GMC_iden}

We identify GMCs using our ALMA \textsuperscript{12}CO(2-1) data and adapted versions of the {\tt CPROPS} algorithms \citep{Rosolowsky2006PASP..118..590R, Liu2021MNRAS.505.4048L}. {\tt CPROPS} picks out `islands' of emission in the position-position-velocity data cube, beginning from pixels brighter than $2$ times the root mean square (rms) noise and growing to neighbouring pixels brighter than $1.5$~rms. The structures identified are also required to be larger than an area of $24$~pixels (twice the synthesised beam area), so that they are physically meaningful, and they are used to create a moment-0 map and identify GMCs. These thresholds are selected to retain almost all diffuse gas while limiting the inclusion of noise. Within these islands, {\tt CPROPS} identifies local maxima and their uniquely associated spaxels as candidate GMCs. Within this pool of GMC candidates, additional criteria are set to select physically-meaningful structures as GMCs. These include a minimum area ($20$~spaxels or $\approx1.7$ times the synthesised beam area), minimum velocity width ($2$~km~s$^{-1}$ or one channel), minimum contrast above the brightest merging brightness level with an adjacent local maximum ($\Delta T_{\rm min}=0.65$~K or $1$~rms) and minimum convexity ($0.45$, lower than typical for ETG and LTG GMCs). The convexity threshold is used to avoid GMCs having too many sub-structures, and is explained in detail in our previous works \citep{Liu2021MNRAS.505.4048L, choi2023wisdom, Liang1387inprep}. We note that these criteria are less stringent than those normally used, but using any higher $\Delta T_{\rm min}$ and/or minimum convexity does not allow for the detection of any GMC.

With this set of rather loose parameters, we identify $36$ resolved clouds and $16$ unresolved clouds, illustrated in Figure~\ref{fig:gmc}. For a cloud to be regarded as resolved, its deconvolved radius and velocity dispersion must be larger than the spatial and spectral instrumental resolution, respectively. Compared to other ETGs (e.g.\ NGC~4526, with a synthesised beam size of $\approx18$~pc, \citealt{utomo2015giant}; NGC~4429, with a synthesised beam size of $\approx14$~pc, \citealt{Liu2021MNRAS.505.4048L}), that typically have hundreds of well-identified GMCs at this spatial scale ($\approx30$~pc), NGC~524 has a much smaller number of GMCs, without well-identified boundaries (low $\Delta T_{\rm min}$) and shapes (low minimum convexity). This suggests that the GMCs identified are well mixed into a smooth and diffuse molecular gas disc, in line with the disc' small smoothness index.

We note that the datacube used for GMC identification (with a $0\farcs38\times0\farcs29$ synthesised beam and $2$~km~s$^{-1}$ channels) was also used by \citet{Smith2019MNRAS.485.4359S} for a different purpose. They showed that these ALMA observations actually recover more flux than the IRAM $30$-m telescope single-dish observations of \citet{Young2011MNRAS.414..940Y}, presumably due to the larger primary beam. Considering the inclusion of the short ACA baselines that cover the entire CO disc (smaller than the maximum recoverable scale) and the well-sampled $uv$ plane, it is therefore likely that (almost) all the flux has been recovered. 

Another caveat is that our GMC identification is constrained by the spatial resolution of our ALMA observations. NGC~524 was observed with a synthesised beam size of $\approx37$~pc, larger than those of some recent ETG studies \citep[e.g.][]{utomo2015giant,Liu2021MNRAS.505.4048L}. It is therefore likely that higher-resolution observations would lead to more robust GMC detections. Nevertheless, the synthesised beam size of our ALMA observations is comparable to that of some GMC surveys of nearby galaxies that successfully identified hundreds of GMCs with radii of $\approx30$ -- $100$~pc in each galaxy \citep[e.g.][]{2021MNRAS.502.1218Roso}. Thus, the paucity of robustly-identified GMCs at this spatial resolution indicates that the small-scale molecular gas morphology of NGC~524 deviates from the norm of nearby galaxies. In Sections~\ref{dis:sim} and \ref{dis:shear}, we will attempt to estimate the sizes of NGC~524's GMCs.

\subsubsection{GMC properties}
\label{GMC_prop}

We measure cloud properties including radius ($R_{\rm c}$), molecular gas mass ($M_{\rm gas}$) and mass surface density ($\Sigma_{\rm gas}\equiv M_{\rm gas}/\pi R_{\rm c}^2$) and velocity dispersion ($\sigma_{\rm obs,los}$) following \citet{Liu2021MNRAS.505.4048L} and references therein (particularly \citealt{Rosolowsky2006PASP..118..590R}). Due to the small size of the GMC sample of NGC~524, the GMC properties do not follow well-defined Gaussian distributions. The GMC radii range from $\approx30$ to $\approx140$~pc with a mean of $\approx60$~pc. We note that the GMC radii are constrained by the limited spatial resolution of our observations ($\approx35$~pc). In Section~\ref{dis:sim}, we quantify this effect by performing a comparison to GMCs identified in a simulated galaxy with a spatial resolution of $5$~pc within the molecular gas reservoir. The means of the other properties are $\langle M_{\rm gas}\rangle\approx6.3\times10^5$~M$_\odot$, $\langle\Sigma_{\rm gas}\rangle\approx126$~${\rm M_\odot~pc^{-2}}$ and $\langle\sigma_{\rm obs,los}\rangle\approx8.2$~km~s$^{-1}$. Compared to other ETGs with GMC catalogues, the clouds of NGC~524 are larger, have higher velocity dispersions and have lower molecular gas mass surface densities. 

Using these cloud properties, we can derive the virial parameters of the GMCs. The virial parameter of a cloud is generically defined as the ratio of its virial mass to its luminous mass, where the virial mass is defined as the dynamical mass at equilibrium. In practice, we measure here the ratio $\alpha_{\rm obs,vir}$ of a cloud's virial mass to its molecular gas mass ($M_{\rm gas}$), where the virial mass is calculated from the measured cloud properties:
\begin{equation}
  M_{\rm obs,vir}\equiv\frac{5\sigma_{\rm obs,los}^2R_{\rm c}}{G}
\end{equation}
\citep{MacLaren1988ApJ...333..821M}, where $G$ is the gravitational constant. 
In the middle panel of Figure~\ref{fig:gmc}, we show $M_{\rm obs,vir}$ as a function of $M_{\rm gas}$, with the distribution of $\alpha_{\rm obs,vir}$ shown in an inset. The identified GMCs have a very high mean virial parameter $\langle\alpha_{\rm obs,vir}\rangle\approx 5.3$. This is consistent with the beam-by-beam analysis  of \citet{williams2023wisdom}, who reported a virial parameter of $6$ at a $60$~pc spatial scale, $8$ at $90$~pc scale and $10$ at $120$~pc scale. A virial parameter significantly higher than $1$ suggests that either the structures are not gravitationally bound and are thus transient, or there are large non-gravitational forces (e.g.\ magnetic fields and/or external pressure) keeping the structures together. It has also been shown by \citet{williams2023wisdom} that the virial parameters do not vary significantly as a function of beam size, but are usually slightly smaller with smaller beam sizes. Thus, it is likely that the virial parameters of the NGC~524 GMCs will remain higher than $1$ even if the spatial resolution of our observations is improved. In Section~\ref{dis:sim}, we further confirm this with a simulated galaxy.

Following the Larson relations \citep{1981MNRAS.194..809Larson}, the bottom panel of Figure~\ref{fig:gmc} shows the size -- linewidth relation of the NGC~524 GMCs, which allows to probe GMC turbulence. The black line is the best-fitting power-law relation between $\sigma_{\rm obs,los}$ and $R_{\rm c}$, with a power-law index of $0.67\pm0.28$ and a Spearman rank correlation coefficient of $0.61$. The size -- linewidth relations of the MW (disc and centre) GMCs and  ETG GMCs (NGC~4429, \citealt{Liu2021MNRAS.505.4048L}; NGC~4526, \citealt{utomo2015giant}; NGC~1387, \citealt{Liang1387inprep}) are shown with coloured contours encompassing $68\%$ of the distributions of the data points. Although constrained by the limited spatial resolution of our \textsuperscript{12}CO(2-1) observations, the NGC~524 GMCs are larger than all other GMCs. At fixed similar cloud size, the linewidths of the NGC~524 GMCs are larger than those of the MW disc and NGC~1387 GMCs but smaller than those of the MW centre and NGC~4526 GMCs (similar to those of the NGC~4429 GMCs), although the overlap in size is limited. The power-law index of the size --linewidth relation of NGC~524 is similar to that found in the MW ($0.5$ -- $0.7$; \citealt{solomon1987,kauffmann2017}) and nearby glaxies ($0.4$ -- $0.8$; \citealt{rosolowsky2003,rosolowsky2007,bolatto2008,wong2011LMC}). It is neither particularly shallow nor steep compared to the size -- linewidth relations measured in the centres of other ETGs (e.g.\ no correlation in NGC~4526, \citealt{utomo2015giant}; $0.82\pm0.13$ in NGC~4429, \citealt{Liu2021MNRAS.505.4048L}; $0.29\pm0.11$ in NGC~1387, \citealt{Liang1387inprep}).

\begin{figure}
   \centering
   \includegraphics[width=0.47\textwidth]{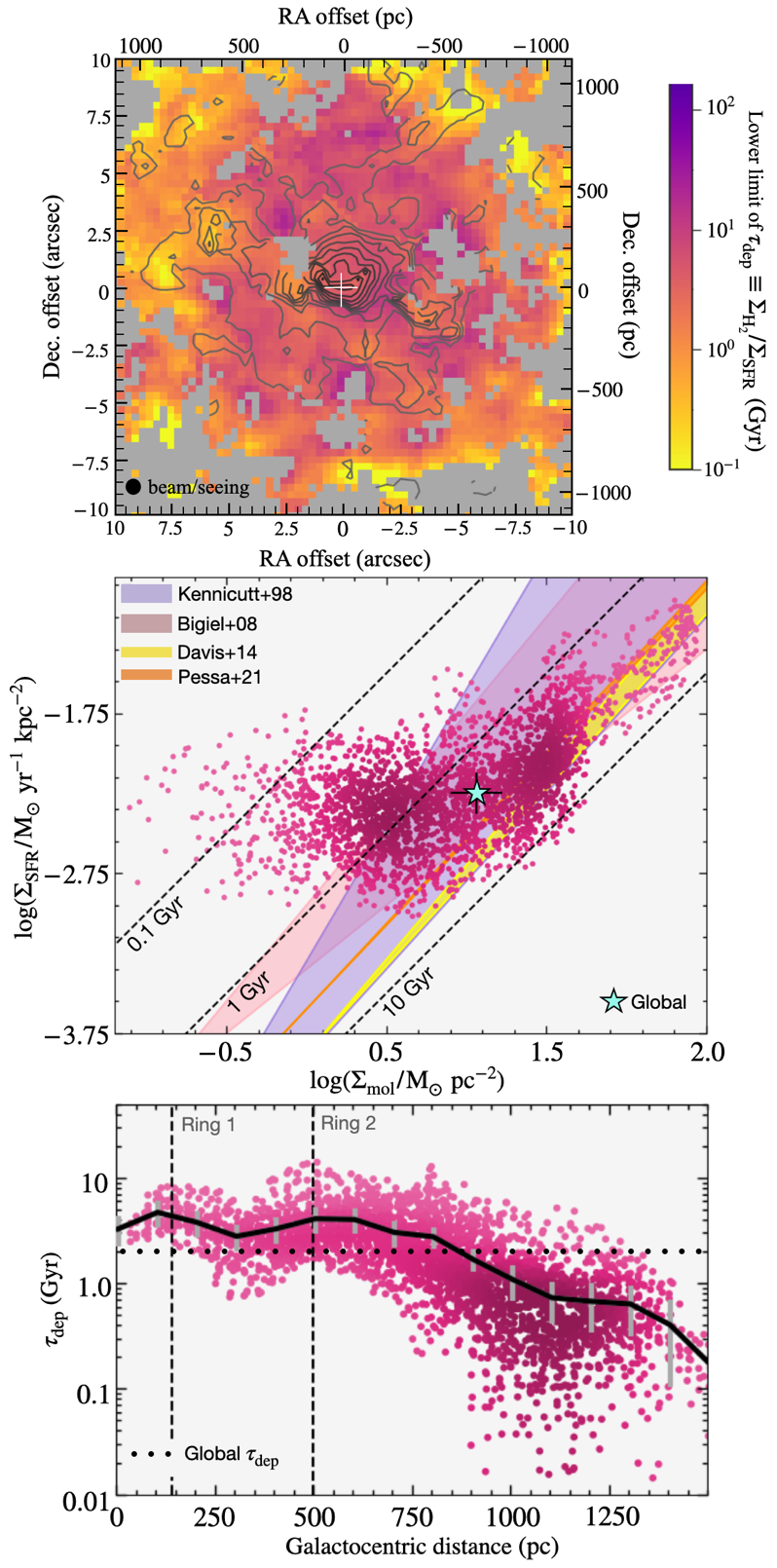}
   \vspace{-0.2cm}
   \caption{\label{fig:sfe} {\it Top:} Depletion time map of NGC~524, for all spaxels with CO and H$\alpha$ surface brightnesses at least three times above the noise level. The extinction-corrected H$\alpha$ surface brightness is overlaid as gray to black contours (from the $3\sigma$ level to the maximum). The centre of the galaxy is marked as a white cross. The synthesised beam/seeing is shown in the bottom-left corner as a solid black circle ($1\farcs1\times1\farcs1$ or $\approx125\times125$~pc$^2$). 
   {\it Middle:} $\Sigma_{\rm SFR}$ -- $\Sigma_{\rm mol}$ relation. Red data points, shaded by their density, satisfy the same criteria as the spaxels in the top panel. The cyan star shows the global $\Sigma_{\rm SFR}$ and $\Sigma_{\rm mol}$. The purple, pink, yellow and orange regions represent the power-law relations of \citet{kennicutt1998global}, \citet{bigiel2008star}, \citet{Davis2014MNRAS.444.3427D} and \citet{2021A&A...650A.134Pessa}, respectively. The black dashed diagonal lines are lines of equal depletion times (labelled). 
   {\it Bottom:} Depletion time as a function of galactocentric distance. Red data points, shaded by their density, satisfy the same criteria as the spaxels in the top panel. The black curve shows the depletion times measured within annuli of $150$~pc width, with the uncertainties shown as grey vertical bars. The black dashed vertical lines indicate the positions of the two molecular gas rings. The black dotted horizontal line indicates the global depletion time. The global molecular gas depletion time of NGC~524 is $\approx2$~Gyr, but this increases to $\approx5.2$~Gyr near the galaxy centre. We note that all depletion times are really lower limits, as the H$\alpha$ emission may not arise purely from SF.}
\end{figure}


\subsection{Star formation efficiency}
\label{SFE}

Taking the ratio of the molecular gas mass of a region and its SFR yields a measurement of the region's depletion time ($\tau_{\rm dep}$), that is the time it would take for the region's gas reservoir to be depleted by star formation should star formation continue at its current rate uninterrupted. The depletion time is the inverse of the SFE. Combining the molecular gas mass surface density map generated in Section~\ref{data:CO} with the SFR surface density map generated in Section~\ref{data:Halpha}, we obtain a spatially-resolved map of depletion time, shown in the top panel of Figure~\ref{fig:sfe}. We include all spaxels for which both the molecular gas and the ionised gas emission have $S/N>3$. At the spatial resolution of the SITELLE data ($\approx125$~pc), both the molecular gas and the ionised gas extend to and cover almost entirely the central $1$~kpc in galactocentric distance. We overlay on the $\tau_{\rm dep}$ map the extinction-corrected H$\alpha$ surface brightness map, to illustrate that regions with smaller $\tau_{\rm dep}$ (yellow to light orange colors) are often co-spatial with H$\alpha$ emission peaks, especially in the outskirts of the galaxy. 

We show the spatially-resolved relation between $\Sigma_{\rm SFR}$ and $\Sigma_{\rm mol}$ in the middle panel of Figure~\ref{fig:sfe} (shaded by the density of data points), and compare the data to the original $\Sigma_{\rm SFR}$ -- $\Sigma_{\rm mol}$ power-law relations of \citeauthor{kennicutt1998global} (\citeyear{kennicutt1998global}; purple region), \citeauthor{bigiel2008star} (\citeyear{bigiel2008star}; pink region), \citeauthor{Davis2014MNRAS.444.3427D} (\citeyear{Davis2014MNRAS.444.3427D}; yellow region) for ETGs and \citeauthor{2021A&A...650A.134Pessa} (\citeyear{2021A&A...650A.134Pessa}; orange region) for nearby star-forming galaxies. 
The global $\tau_{\rm dep}$, calculated as the ratio of the total molecular gas mass over the total SFR within the $20\arcsec\times20\arcsec$ ($\approx2.26\times2.26$~kpc$^2$) region tightly encompassing the molecular gas disc (see Figure~\ref{fig:data} and Sections~\ref{data:CO} and \ref{data:Halpha}), is shown as a cyan star. This global $\tau_{\rm dep}$ is marginally longer than that expected from \citet{kennicutt1998global} and \citet{bigiel2008star} but shorter than that expected from \citet{Davis2014MNRAS.444.3427D} and \citet{2021A&A...650A.134Pessa}. The distribution of the NGC~524 data points covers a wide region in the parameter space of SFR and molecular gas mass surface densities. 
At the high end of $\Sigma_{\rm mol}$, the distribution of depletion time spans $\approx1$ to $\approx10$~Gyr. At the low end of $\Sigma_{\rm mol}$, a tail of depletion times spans $\approx0.1$ (and even less) to $\approx1$~Gyr, and $\Sigma_{\rm SFR}$ remains approximately constant independently of $\Sigma_{\rm mol}$.

The galactocentric distance profile of the depletion time is shown in the bottom panel of Figure~\ref{fig:sfe} (shaded by the density of data points) and reveals a gradually increasing $\tau_{\rm dep}$ with decreasing galactocentric distance. The global depletion time of $2.0$~Gyr is marked in the figure as a black dotted horizontal line. Although the global $\tau_{\rm dep}$ is similar to that of nearby star-forming galaxies, $\tau_{\rm dep}$ is higher than the mean by a factor of $3$ -- $4$ within a galactocentric distance of $\approx900$~pc, and it approaches its maximum $\tau_{\rm dep,max}=5.2$~Gyr near the centre of the galaxy. The two molecular gas rings of NGC~524 (see Section~\ref{data:CO}) are also marked in the panel, and $\tau_{\rm dep}$ has local maxima at the locations of these rings (due to the overdensity of molecular gas but the roughly constant ionised-gas surface brightness). 

We note again however the major bias of our analyses, that all H$\alpha$ emission is included as a SF tracer. We attempted to separate the different potential ionisation mechanisms using standard emission-line ratio diagnostics \citep[e.g.][]{1981PASP...93....5BPT}, but the diagnostics indicate that all the ionised gas (i.e.\ all spaxels/data points) is ionised by old stellar populations, and no emission is characteristic of star-forming regions (see Appendix~\ref{app_Halpha}). Therefore, the SFRs calculated here are really upper limits. Conversely, the depletion times are really lower limits. Of course, this strongly suggests that H$\alpha$ emission arising from SF is very limited, which strengthens our argument that SF is suppressed in NGC~524.

Another parameter that may affect the measurements of $\tau_{\rm dep}$ is $X_{\rm CO(1-0)}$. In this work, we adopted the MW $X_{\rm CO(1-0)}$, that has also been applied to other nearby galaxies \citep[e.g.][]{sun2018cloud,Liu2021MNRAS.505.4048L}. Recent spatially-resolved investigations of nearby galaxies have however uncovered that $X_{\rm CO(1-0)}$ can be smaller in galaxy centres than in galaxy discs \citep{2013ApJ...777....5Sandrom,2022ApJ...925...72Teng}. Such a radial dependence of $X_{\rm CO(1-0)}$ could therefore (partially) explain the trend of increasing $\tau_{\rm dep}$ with decreasing galactocentric radius. However, due to a lack of $X_{\rm CO(1-0)}$ measurements in ETGs, whether lower $X_{\rm CO(1-0)}$ are the result of the bulge-disc transition or whether they are similar in ETGs with pure spherical potentials is currently unknown.

\begin{figure*}
   \centering
   \includegraphics[width=0.95\textwidth]{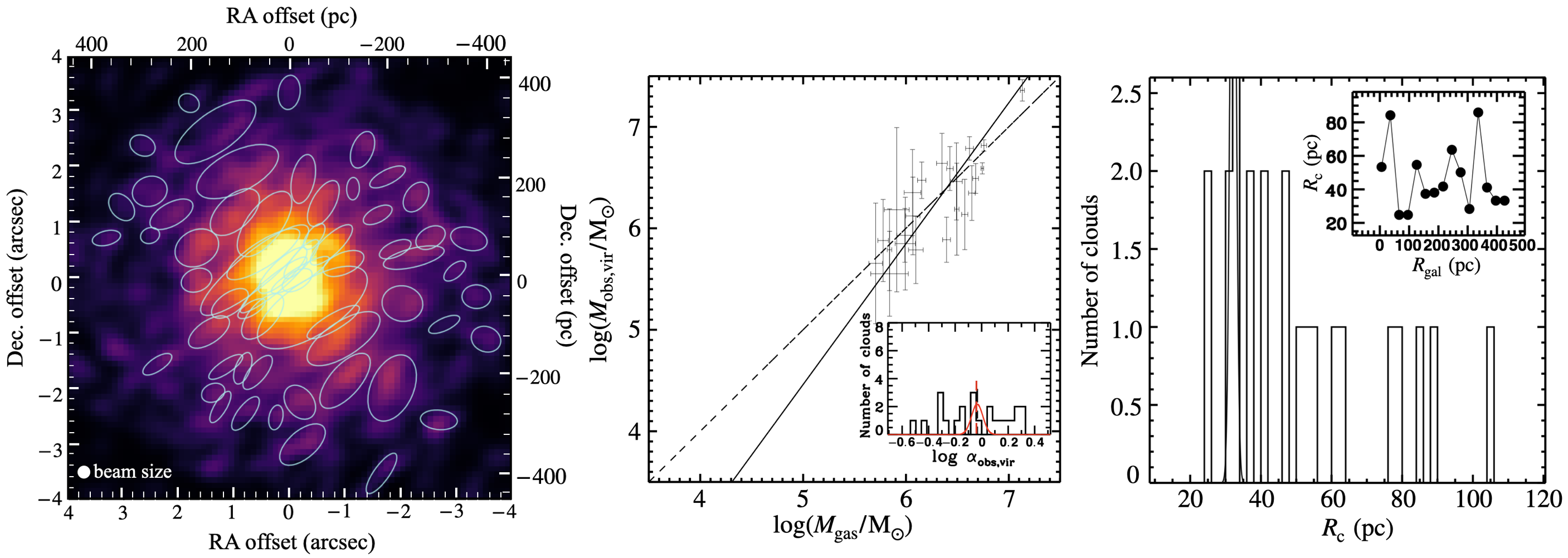}
   \includegraphics[width=0.95\textwidth]{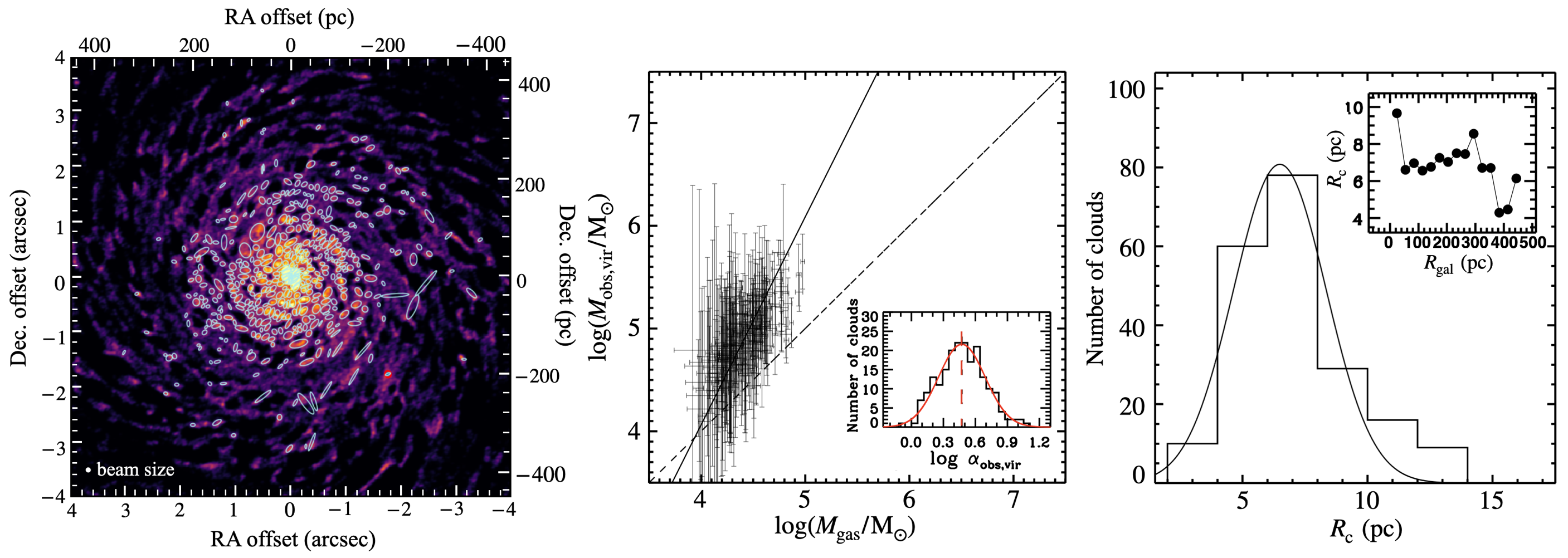}
   \caption{\label{fig:sim} Comparison of the CO data cubes generated at spatial resolutions of $30$ ({\it top row}, matching the resolution of our \textsuperscript{12}CO(2-1) observations) and $5$ ({\it bottom row}, highest resolution possible) pc from a simulated galaxy with properties similar to those of NGC~524. Many more GMCs are identified at the better spatial resolution. {\it Left:} identified GMCs overlaid on the simulated molecular gas moment-0 map. The synthesised beam is shown in the bottom-left corner as a solid white ellipse ($\approx0\farcs27\times0\farcs27$ or $30\times30$~pc$^2$ for the low-resolution simulation; $\approx0\farcs045\times0\farcs045$ or $5\times5$~pc$^2$ for the high-resolution simulation). {\it Middle:} Correlation between virial mass and molecular gas mass for the identified GMCs. The black solid line shows the best-ﬁtting power-law relation, while the black dashed diagonal line indicates the $1:1$ relation. The inset shows the distribution of $\log(\alpha_{\rm obs,vir})$ (black histogram) with a log-normal ﬁt overlaid (red solid line). The red dashed line indicates the mean of the log-normal ﬁt. {\it Right:} distribution of cloud radius (black histogram) with a Gaussian fit overlaid (black solid line). The inset shows the mean cloud radius as a function of galactocentric distance.
   }
\end{figure*}


\section{Comparison with a simulated ETG}
\label{dis:sim}

As our ability to study GMC properties is constrained by the spatial resolution of our ALMA \textsuperscript{12}CO(2-1) observations, we compare our results to those obtained from analogous mock observations of the CO emission from a simulated ETG. The initial conditions and properties of the simulation are described in detail in Jeffreson et al.\ (in prep.), but we provide a brief description here. 

The simulated galaxy is part of a suite of isolated galaxy simulations of ETGs, with initial conditions based on ATLAS$^{\rm 3D}$ \citep{2011MNRAS.413..813C_ATLAS3DI} and MASSIVE \citep{2014ApJ...795..158Massive} measurements of ETG properties. We consider here only the simulated galaxy of total stellar mass $M_\star=10^{11}$~M$_\odot$, which most closely resembles NGC~524. The initial conditions were generated with {\tt MakeNewDisk} \citep{springel2005modelling} and consist of a \citet{hernquist1990analytical} dark matter halo (total halo mass of $10^{13}$~M$_\odot$) and an initial stellar component split into an exponential disc (of mass $M_{\rm disc}=2\times10^{10}$~M$_\odot$) and a \citet{hernquist1990analytical} bulge (of mass $M_{\rm bulge}=8\times10^{10}$~M$_\odot$ and stellar half-mass radius $2.8$~kpc). The initial cold gas fraction is $\log(M_{\rm mol}/M_{\rm \star})=-2.8$ and only a negligible amount of gas is converted into stars across the $400$~Myr runtime. Gas particles have a mass resolution of $859$~M$_\odot$ and employ an adaptive gravitational softening with a minimum softening length of $3$~pc. Star (dark matter) particles have a mass resolution of $5\times10^3$~M$_\odot$ ($1.35\times10^5$~M$_\odot$) and a fixed gravitational softening of $3$~pc ($280$~pc).

The simulations were run with the moving mesh code {\tt Arepo}, including the non-equilibrium network for hydrogen, carbon and oxygen chemistry described in \citet{nelson1997dynamics} and \citet{glover2007simulating}, coupled to the atomic and molecular cooling function of \citet{glover2010modelling}, which self-consistently allows the formation of a multi-phase ISM that includes a cold molecular phase. Stars can form in cold and dense gas with a SFE that depends on the turbulent state of the gas via the virial parameter \citep{2016ApJ...822...11Padoan}, using the implementation of \citet{gensior2020heart}. Stellar feedback is included via energy and momentum injections from supernovae as well as pre-supernova feedback from \ion{H}{II} regions, following the prescription of \citet{jeffreson2021scaling}.

Once it has reached a state of dynamical equilibrium, the simulated galaxy has properties (stellar mass $\log(M_\star/{\rm M_\odot})=11.03$, molecular hydrogen mass $\log(M_{\rm H_2}/{\rm M_\odot})=7.55$ and SFR $\log({\rm SFR}/{\rm M_\odot~yr^{-1}})=-1.44$) that are comparable to those of NGC~524 ($\log(M_\star/{\rm M_\odot})=11.4$, $\log(M_{\rm H_2}/{\rm M_\odot})=7.95$ and $\log({\rm SFR}/{\rm M_\odot~yr^{-1}})=-0.56$). The SFR of the simulated galaxy is formally much lower than that of NGC~524, but it is measured very differently.  Specifically, the SFR of the simulated galaxy is obtained by summing up the initial masses of all star particles formed within the past $5$~Myr and then dividing by the $5$~Myr timescale. While the $5$~Myr timescale is chosen in analogy to the timescales/lifetime of stars traced by H$\alpha$, it is only an approximation of the flux-based SFR measurement for NGC~524.

Based on the molecular gas mass distribution and kinematics of the simulated galaxy, we create mock data cubes that incorporate observational effects such as beam smearing and velocity binning, using the {\tt Kinematic Molecular Simulation} ({\tt KinMS}) tool of \citet{Davis2013MNRAS.429..534D}. The input parameters of {\tt KinMS} include the positions, circular velocity curve and CO luminosity (generated using {\tt DESPOTIC}) of the simulation particles, as well as the distance, inclination and other observational properties of NGC~524. {\tt DESPOTIC} is a software library that calculates the energy and spectra of optically-thick interstellar clouds \citep{Krumholz2014despotic}.

To better understand the GMC sizes and properties under this kind of gravitational potential and gas fraction, we generated two data cubes from the simulated galaxy: one matching the spatial resolution of our observations ($30$~pc), the other at the highest resolution possible ($5$~pc). We also matched the $S/N$ of these simulated data cubes to those of our observational data of NGC~524. To achieve this, we created random noise following a Gaussian distribution at each position-position-velocity pixel, and scaled this noise so that the $S/N$ (i.e.\ the ratio of the mean CO surface brightness of the galaxy disc to the standard deviation of the noise pixels) is the same as that of the NGC~524 data cube.

We then used {\tt CPROPS} \citep{Rosolowsky2006PASP..118..590R, Liu2021MNRAS.505.4048L} to identify and measure the properties of the  GMCs in both data cubes. To ensure a fair comparison to the observations, we used the same GMC selection criteria as stated in Section~\ref{GMC_iden}: minimum area limit of $20$ spaxels, minimum channel width of $2$~km~s$^{-1}$, $\Delta T_{\rm min}=0.65$~K and minimum convexity of $0.45$. The GMCs identified are overlaid on the moment-0 map of the associated data cube in Figure~\ref{fig:sim}. The differences between the two data cubes (and identified GMCs) are significant. At low spatial resolution, matching our observations, the simulated molecular gas disc appears similarly smooth, with just a few large GMCs identified. At high resolution, tightly wound spiral features are revealed, that enable the detection of hundreds more GMCs. With the high resolution data, we are able to constrain the average GMC size to be $\approx6$~pc.  
The average virial parameter is $\approx3$, higher than typical for nearby spirals ($\approx1$) but lower than that of NGC~524 ($\approx5$), indicating that like the observed structures, none of the resolved, simulated GMCs are gravitationally bound. Although the stellar, molecular gas and SF properties of this simulated galaxy are all similar to those of NGC~524, this simulation does not contain the molecular gas rings detected.

Comparing these two data cubes from the same simulated galaxy to our observational data cube of NGC~524, we conclude that it is highly likely that the identification (and thus properties) of the GMCs of NGC~524 is limited by the spatial resolution of the observations. For both NGC~524 and the simulated galaxy at low resolution, the molecular gas disc appears smooth, enabling the detection of only a few large GMCs of sizes $\approx30$~pc, similar to the spatial resolution. However, for the simulated galaxy at high resolution, hundreds of GMCs are detected, with sizes $\approx6$~pc. Interestingly, this is much smaller than the GMCs detected in a simulated Milky Way-like galaxy ($\approx25$~pc; see \citealt{2020MNRAS.498..385Jeffreson}), as well as the typical GMC sizes of nearby galaxies ($10$ -- $30$~pc), although for nearby galaxies the GMC sizes are always constrained by the spatial resolutions of the observations.


\section{Discussion}
\label{discussion}

With our ALMA and SITELLE observations, we have established that NGC~524 has a smooth molecular gas disc and diffuse ionised gas. We are only able to identify a few dozens GMCs, that all have very high virial parameters, suggesting that these GMCs are unbound and unlikely to form stars. By comparing with a similar simulated ETG, we showed that the real GMCs could in fact be much smaller than observed. The SFE upper limits are low (high $\tau_{\rm dep}$ lower limits) compared to those of nearby star-forming galaxies, particularly in the central $1$~kpc galactocentric distance region. These low SFEs are confirmed by $22~\mu$m + FUV SFRs, as discussed in Section~\ref{SFE}. Here, we discuss what is shaping the molecular gas properties and SFE of NGC~524.


\subsection{Strong shear\textit{}}
\label{dis:shear}

A unique feature of NGC~524 is the very high rotation velocity of the molecular gas ($\gtrsim350$~km~s$^{-1}$). The associated high stellar mass surface density creates a deep and steep gravitational potential well, and in turn strong shear affecting the molecular gas. Here we adopt the circular velocity curve derived by \citet{Smith2019MNRAS.485.4359S} through modelling of the stellar mass distribution and molecular gas kinematics. The circular velocity increases rapidly from the center of the galaxy to reach $\approx300$~km~s$^{-1}$ at a galactocentric distance of only $\approx100$~pc, and then gradually rises to $\gtrsim430$~km~s$^{-1}$ at a galactocentric distance of $\approx1$~kpc. 

\subsubsection{Cloud mass surface density}
\label{dis:shear_density}

One way to assess the strength of shear is to evaluate the equilibrium between a cloud’s self-gravitational energy and its kinetic energy associated with shear motions. For a cloud to be gravitationally bound, the contribution of external gravity to the cloud’s energy budget must not exceed the cloud’s self-gravitational energy. Assuming that the cloud's external energy is from shear motions due to the circular rotation of the galactic disc, this requirement can be expressed as $4A_0^2R_{\rm c}/3\pi G\Sigma_{\rm gas}\leq1$ (see Equations~32 and 48 of \citealt{Liu2021MNRAS.505.4048L}, where $A_0\equiv-\frac{R}{2} \frac{d\Omega(R)}{dR}|_{R=R_0}$ is Oort's constant $A$ evaluated at the cloud centre $R_0$, $R$ is the galactocentric distance and $\Omega(R)\equiv V_{\rm circ}(R)/R$ is the angular frequency of circular motion, where $V_{\rm circ}(R)$ is the circular velocity curve). This implies that for a given cloud radius $R_{\rm c}$, there is a minimum surface density $\Sigma_{\rm shear}$ required for the cloud to remain marginally bound (see also Equation~55 of \citealt{Liu2021MNRAS.505.4048L}):
\begin{equation}
   \Sigma_{\rm shear}\approx\frac{4A_0^2R_{\rm c}}{3\pi G}\,\,\,.
\end{equation}

\begin{figure*}
   \centering
   \includegraphics[width=0.98\textwidth]{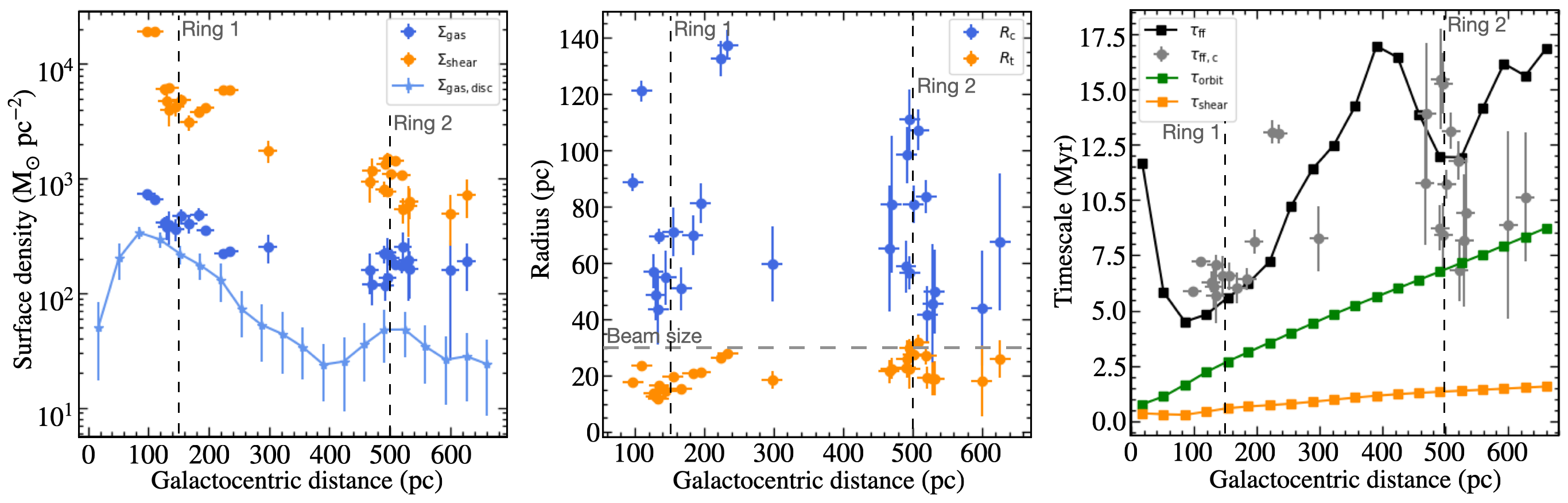}
   \caption{\label{fig:shear} Key surface density, length and time scales as a function of galactocentric distance, illustrating that shear is the dominant mechanism shaping the identified NGC~524 clouds. {\it Left:} galactocentric distance profiles of the molecular gas mass surface densities measured for each identified GMC ($\Sigma_{\rm gas}$; dark blue data points), required for each identified GMC to withstand shear ($\Sigma_{\rm shear}$; orange data points), and azimuthally averaged over the galaxy disc ($\Sigma_{\rm gas,disc}$; light blue data points and curve). {\it Middle:} galactocentric distance profiles of the radii measured for each identified GMC ($R_{\rm c}$; dark blue data points) and required for each identified GMC to be destroyed by shear ($R_{\rm t}$; orange data points). The pale grey dashed horizontal line indicates the synthesised beam size. {\it Right:} galactocentric distance profiles of the molecular gas disc free-fall ($\tau_{\rm ff}$; black data points and curve), each identified GMC free-fall ($\tau_{\rm ff,c}$; gray data points), orbital ($\tau_{\rm orbit}$; green data points and curve) and shear ($\tau_{\rm shear}$; orange data points and curve) timescales. In all panels, the dark grey dashed vertical lines indicate the positions of the two molecular gas rings.}
\end{figure*}

A comparison of $\Sigma_{\rm shear}$, $\Sigma_{\rm gas}$ and the azimuthally-averaged molecular gas mass surface density of the disc ($\Sigma_{\rm gas,disc}$) of NGC~524 is shown in the left panel of Figure~\ref{fig:shear}. As by construction the GMCs identified are over-densities of the molecular gas disc, it is expected that $\Sigma_{\rm gas}$ is larger than $\Sigma_{\rm gas,disc}$. However, across the galaxy disc $\Sigma_{\rm shear}$ is also systematically larger than $\Sigma_{\rm gas}$, and the difference gradually increases toward the galaxy centre, where $\Sigma_{\rm shear}$ is almost one order of magnitude higher than $\Sigma_{\rm gas}$. This indicates that the energy associated with shear motions significantly exceeds the self-gravitational energy of the identified GMCs, strongly suggesting that the NGC~524 GMCs are not gravitationally bound.

\subsubsection{Cloud radius}
\label{dis:shear_radius}

Similarly, we use Equation~52 of \citet{Liu2021MNRAS.505.4048L} to estimate the tidal radius of each GMC:
\begin{equation}
   R_{\rm t}=\left(\frac{G}{2A_0^2}\right)^{1/3}M_{\rm c}^{1/3}\,\,\,,
\end{equation}
where $M_{\rm c}$ is the mass of the cloud, taken here to be $M_{\rm gas}$. This tidal radius is defined as the distance from the GMC centre where the shear velocity due to differential galactic circular rotation is equal to the escape velocity of the cloud \citep{1991ApJ...378..565Gammie, 2000ApJ...536..173Tan}. We compare $R_{\rm t}$ to our measured $R_{\rm c}$ for the identified GMCs of NGC~524 in the middle panel of Figure~\ref{fig:shear}. The cloud radii are systematically much larger than the tidal radii, suggesting again that the identified clouds are gravitationally unbound.

This analysis also suggests that gravitationally-bound clouds in NGC~524 cannot have radii larger than $\approx30$~pc, which is about the synthesised beam size of our ALMA observations. This may be the reason we cannot confidently identify GMCs in NGC~524 (see Section~\ref{GMC_iden}). Compared to the clouds of other ETGs \citep[][]{utomo2015giant, Liu2021MNRAS.505.4048L, Liang1387inprep}, the mean tidal radius of the identified NGC~524 clouds is the smallest, and the difference between $R_{\rm c}$ and $R_{\rm t}$ is the largest. In such a system with a smooth molecular gas disc and strong shear, the clouds formed will likely be very small, with a radius of only a few parsecs. The spatial resolution of our observation would have to be much higher to be able to distinguish such clouds. Using the simulated galaxy discussed in Section~\ref{dis:sim} at very high resolution, we do detect GMCs with a mean radius of $\approx6$~pc, in line with our expectations. Therefore, based on our calculations and the comparison with a simulated galaxy, we estimate that a spatial resolution better than $15$~pc, preferably better than $6$~pc, is necessary to determine if there are any gravitationally-bound GMCs.

\subsubsection{Timescales}
\label{dis:shear_lifetime}

We compare each cloud's shear and orbital timescales ($\tau_{\rm shear}$ and $\tau_{\rm orbit}$) with its free-fall timescale ($\tau_{\rm ff}$) to establish which is more important: shear or self gravity. The mechanism with the shorter timescale must dictate the structure and dynamics of the cloud. The orbital timescale describes the time taken by a GMC to complete one circular orbit. As a cloud moves along its orbit and experiences shear, it becomes larger and less bound over the shear timescale. The orbital and shear timescales are intrinsically related:
\begin{equation}
   \tau_{\rm orbit}\equiv\tau_{\rm orbit}(R)=2\pi/\Omega(R)
\end{equation}
and
\begin{equation}
   \begin{split}
      \tau_{\rm shear}\equiv\tau_{\rm shear}(R) & =\frac{1}{2A(R)} \\
      & \approx t_{\rm orbit}(R)/2\pi\,\,\,,
   \end{split}
\end{equation}
where the last expression is only valid for a flat circular velocity curve. The free-fall time ($\tau_{\rm ff}\equiv\sqrt{3\pi/32G\rho}$, where $\rho$ is the mass volume density) is the timescale over which a molecular gas cloud will collapse under its own gravity. Following \citet{2018ApJ...861L..18Utomo}, we use
\begin{equation}
   \tau_{\rm ff}=\sqrt{\frac{3\pi H}{32G\Sigma_{\rm gas,disc}}}\,\,\,,
\end{equation}
where $H$ is the line-of-sight depth of the molecular gas layer, taken here to be $100$~pc based on the characteristic thickness of the molecular gas layer of the MW and other galaxies \citep{2013ApJ...779...43Pety, 2014AJ....148..127Yim, 2015ARA&A..53..583Heyer}. This allows us to estimate the free-fall time of structures with radii of $\approx100$~pc throughout the molecular gas disc. As a comparison, we also calculate the free-fall time of our identified cloud ($\tau_{\rm ff,c}$) assuming a spherical shape ($\rho=3M_{\rm gas}/4\pi R_{\rm c}^3$).

We compare these timescales in the right panel of Figure~\ref{fig:shear}, showing that $\tau_{\rm shear}$ and $\tau_{\rm orbit}$ are significantly shorter than $\tau_{\rm ff}$ (for structures within the molecular gas disc) and $\tau_{\rm ff,c}$ (for all identified GMCs) across NGC~524. This indicates again that shear is the dominant mechanism regulating the molecular gas behaviour in NGC~524, consistent with the smooth molecular gas disc observed and the long molecular gas depletion times. 


\subsection{Non-circular motions}
\label{dis:noncirc}

As discussed in \citet{Smith2019MNRAS.485.4359S}, there are non-circular molecular gas motions in NGC~524, due to weak bi-symmetric spiral structures revealed by the mean line-of-sight velocity residual map and the velocity dispersion map. The former was generated by subtracting an axisymmetric rotating disc model from the observed velocity field.  
At the same location as the observed spiral pattern, there is also an annular structure with higher velocity dispersions, clearly visible in the observed moment-2 map. This is interpreted as due to beam smearing of both the rotation velocities and the spiral structures. 

The line-of-sight velocities were further separated into azimuthal and radial components by \citet{Smith2019MNRAS.485.4359S}, using the harmonics of the observed velocity field. This decomposition indicates that the non-circular motions can be explained by either a warped disc or an outflow, potentially driven by the AGN. As discussed in Section~\ref{data:target}, there is a compact radio source at the centre of NGC~524, but no clear evidence of nuclear activity from either radio observations or our ionised-gas maps.

Overall, while non-circular motions are present in NGC~524, especially in the central $\approx300$~pc in galactocentric distance, there is no direct link between them and potential radial inflows or outflows that can add turbulence to the GMCs and affect their SFEs. Even if these non-circular motions were associated with inflows/outflows, the residual velocities are only $\approx10\%$ of the rotational velocities, so their amplitudes are limited, especially compared to the dominant shear motions due to the fast-rotating disc.


\subsection{Comparison to other ETGs}
\label{dis:otherETG}

Compared to other ETGs in the WISDOM sample \citep{Davis2022MNRAS.512.1522D}, NGC~524 has almost the highest stellar mass surface density, but there are a few galaxies that appear smoother (lower smoothness index). Interestingly, the ETGs that are smoother (e.g.\ NGC~4429, NGC~1387 and NGC~383) all have well-identified GMCs with sizes $\lesssim30$~pc. The NGC~4429 GMCs have properties that suggest self gravity at best just counterbalances shear \citep{Liu2021MNRAS.505.4048L}. The GMCs of NGC~1387 are more similar to those of the MW, with virialised clouds of sizes comparable to their tidal radii \citep{Liang1387inprep}. 
Although these ETGs have a variety of GMC properties, it is unusual that it is so difficult to identify GMCs in NGC~524, with at best a few GMCs with very high virial parameters. 

Within the WISDOM sample of ETGs, NGC~524 is a rare case that clearly demonstrates the impact of the deep and steep gravitational potentials of spheroids and the associated strong shear. As shown in \citet{williams2023wisdom}, the molecular gas of NGC~524 has high virial parameters that are typical of ETGs, but the turbulent pressures are the lowest, indicating that the GMCs of NGC~524 are not bound by gravity but are also not turbulent. This is in line with what was demonstrated in Section~\ref{dis:shear}: shear is the dominant force shaping the molecular gas properties of NGC~524. Another unique feature of NGC~524 is that it has the smallest molecular gas fraction of the WISDOM ETG sample \citep{Davis2022MNRAS.512.1522D}. This may also offer an explanation for the strong impact of shear, as simulations show that the impact of the potential becomes stronger with lower cold gas fractions \citep{gensior2021elephant}. Although all of the sample ETGs have deep gravitational potential wells created by the centrally-concentrated stellar distributions, NGC~524 has the lowest molecular gas mass surface density within the central $1$~kpc, and hence the weakest molecular cloud self gravity to balance the strong shear from the galaxy rotation. This strong shear does not imply that the GMCs are particularly turbulent, but it can destroy GMCs before they collapse and form stars. This is in line with the observed long depletion times discussed in Section~\ref{SFE}. 


\section{Conclusions}
\label{conclusion}

In this work, we have used ALMA and SITELLE observations to study the molecular and ionised gas of NGC~524. We find that:

\begin{enumerate}
\item Based on our ALMA observations (synthesised beam size of $\approx37$~pc), NGC~524 has a smooth molecular gas disc, with at best a few dozens GMCs that have small molecular gas mass surface densities, high virial parameters ($\langle\alpha_{\rm obs,vir}\rangle\approx5.3$) and no well-defined edge.

\item The ionised-gas emission is not directly associated with star formation. The SFR upper limits inferred from the H$\alpha$ emission imply depletion time lower limits with a global $\tau_{\rm dep}$ of $\approx2$~Gyr and a maximum of $\approx5.2$~Gyr near the galaxy centre.

\item Based on analyses of the identified GMC molecular gas mass surface densities, radii and relevant timescales, shear is the dominant force regulating GMC properties and the resultant SF in NGC~524.

\item The GMC analyses of this work are constrained by the spatial resolution of our observations. It is likely that GMCs in galaxies similar to NGC~524 have sizes of only a few parsecs, a behaviour supported by the comparison of our observations with a simulated galaxy whose properties closely match those of NGC~524. Thus, molecular gas observations with better spatial resolutions are required to unveil the details of the NGC~524 GMCs.

\end{enumerate}

\section*{Acknowledgements}

We thank the anonymous referee for suggestions and comments. This research is based on observations obtained with the SITELLE instrument on the Canada-France-Hawaii Telescope (CFHT) which is operated from the summit of Maunakea, and the Atacama Large Millimeter/submillimeter Array (ALMA) in the Atacama desert.
This paper makes use of the following ALMA data: ADS/JAO.ALMA\#2015.1.00466.S, ADS/JAO.ALMA\#2016.2.00053.S, and ADS/JAO.ALMA\#2017.1.00391.S. ALMA is a partnership of ESO (representing its member states), NSF (USA) and NINS (Japan), together with NRC (Canada), MOST and ASIAA (Taiwan), and KASI (Republic of Korea), in cooperation with the Republic of Chile. The Joint ALMA Observatory is operated by ESO, AUI/NRAO and NAOJ. The National Radio Astronomy Observatory is a facility of the National Science Foundation operated under cooperative agreement by Associated Universities, Inc.

We are grateful to the CFHT and ALMA scheduling, data processing and archive teams. We also wish to acknowledge that the summit of Maunakea is a significant cultural and historic site for the indigenous Hawaiian community, while the the high-altitude plateau Chajnantor on which the ALMA telescope sits is sacred to indigenous Likanantai people. We are most grateful to have the opportunity of observing there. We thank Nicole Ford for the scientific and aesthetic support.

AL, HB, DH, CR, LD and LRN acknowledge funding from the NSERC Discovery Grant and the Canada Research Chairs (CRC) programmes. MB was supported by STFC consolidated grant `Astrophysics at Oxford' ST/H002456/1 and ST/K00106X/1. JG gratefully acknowledges financial support from the Swiss National Science Foundation (grant no CRSII5\_193826). SMRJ is supported by Harvard University through the Institute for Theory and Computation Fellowship. IR acknowledges support from grant ST/S00033X/1 through the UK Science and Technology Facilities Council (STFC). TAD acknowledges support from the UK Science and Technology Facilities Council through grant ST/W000830/1. 

\section*{Data Availability}
The raw data underlying this article are publicly available on the National Radio Astronomy Observatory (programmes 2015.1.00466.S, 2016.2.00053.S, and 2017.1.00391.S) and CFHT archives (programmes 22Bc09 and 20Bc25). All analysed data products are available upon request.

\bibliographystyle{mnras}
\bibliography{ref_v3}

\newpage

\appendix

\section{Ionised-gas maps}
\label{app_Halpha}

\begin{figure*}
  \centering
  \includegraphics[width=0.98\textwidth]{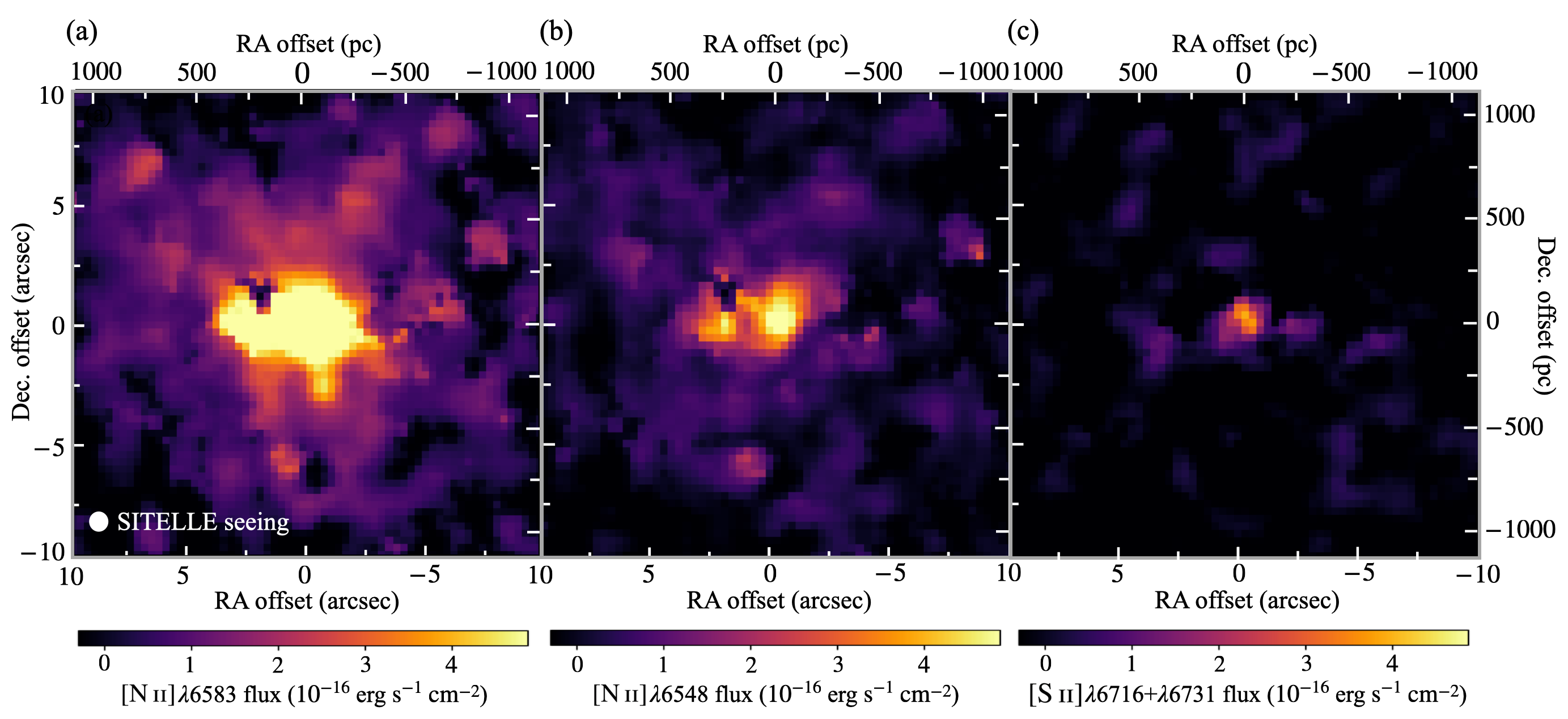}
  \includegraphics[width=0.98\textwidth]{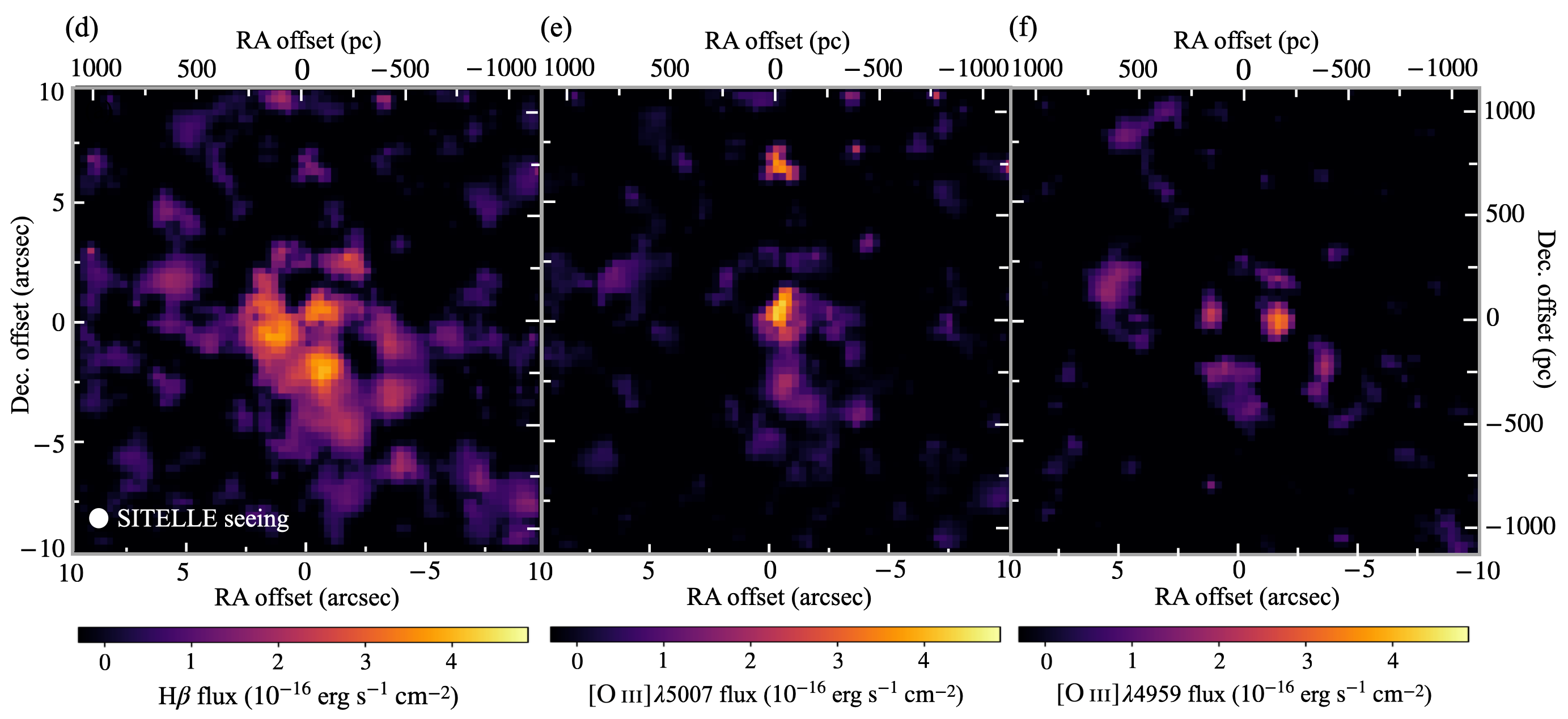}
  \caption{\label{fig:app_1} NGC~524 surface brightness maps of emission lines from the SN3 data cube. (a) [\ion{N}{II}]$\lambda6583$. (b) [\ion{N}{II}]$\lambda6548$. (c)  [\ion{S}{II}]$\lambda6716$~$+$~[\ion{S}{II}]$\lambda6731$. (d) H$\beta$. (e) [\ion{O}{III}]$\lambda5007$. (e) [\ion{O}{III}]$\lambda4959$. The seeing is shown in the bottom-left corner of the leftmost maps as a solid black circle ($1\farcs1\times1\farcs1$ or $\approx125\times125$~pc$^2$).}
\end{figure*}

Figure~\ref{fig:app_1} shows the surface brightness (i.e.\ the flux within one pixel corrected for absorption but not extinction, divided by the pixel area) maps of emission lines from the SN3 data cube, as measured with {\tt ORCS}. We note that the H$\alpha$ map is shown in Figure~2 and is not repeated there.

\begin{figure*}
  \centering
  \includegraphics[width=0.98\textwidth]{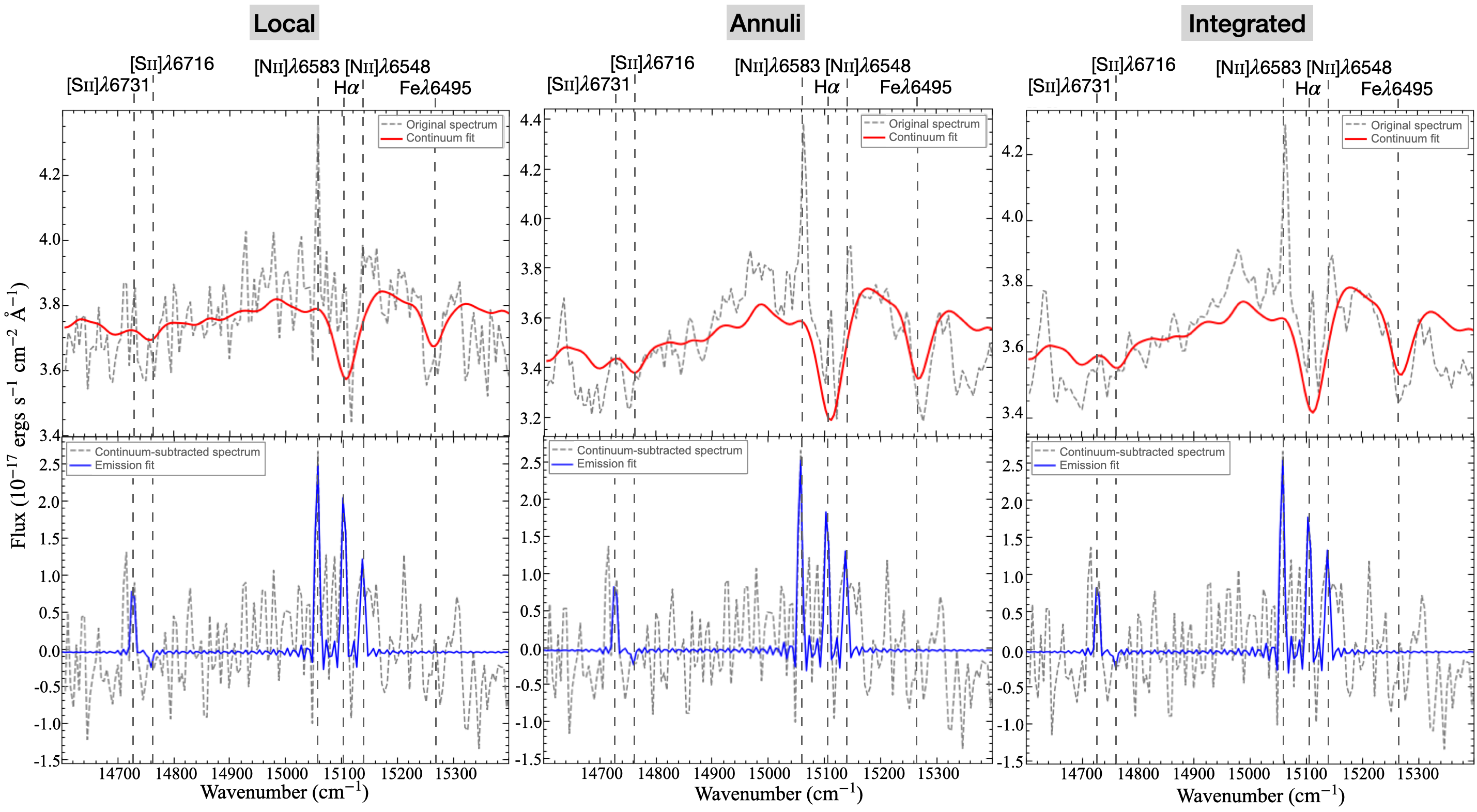}
   \includegraphics[width=0.98\textwidth]{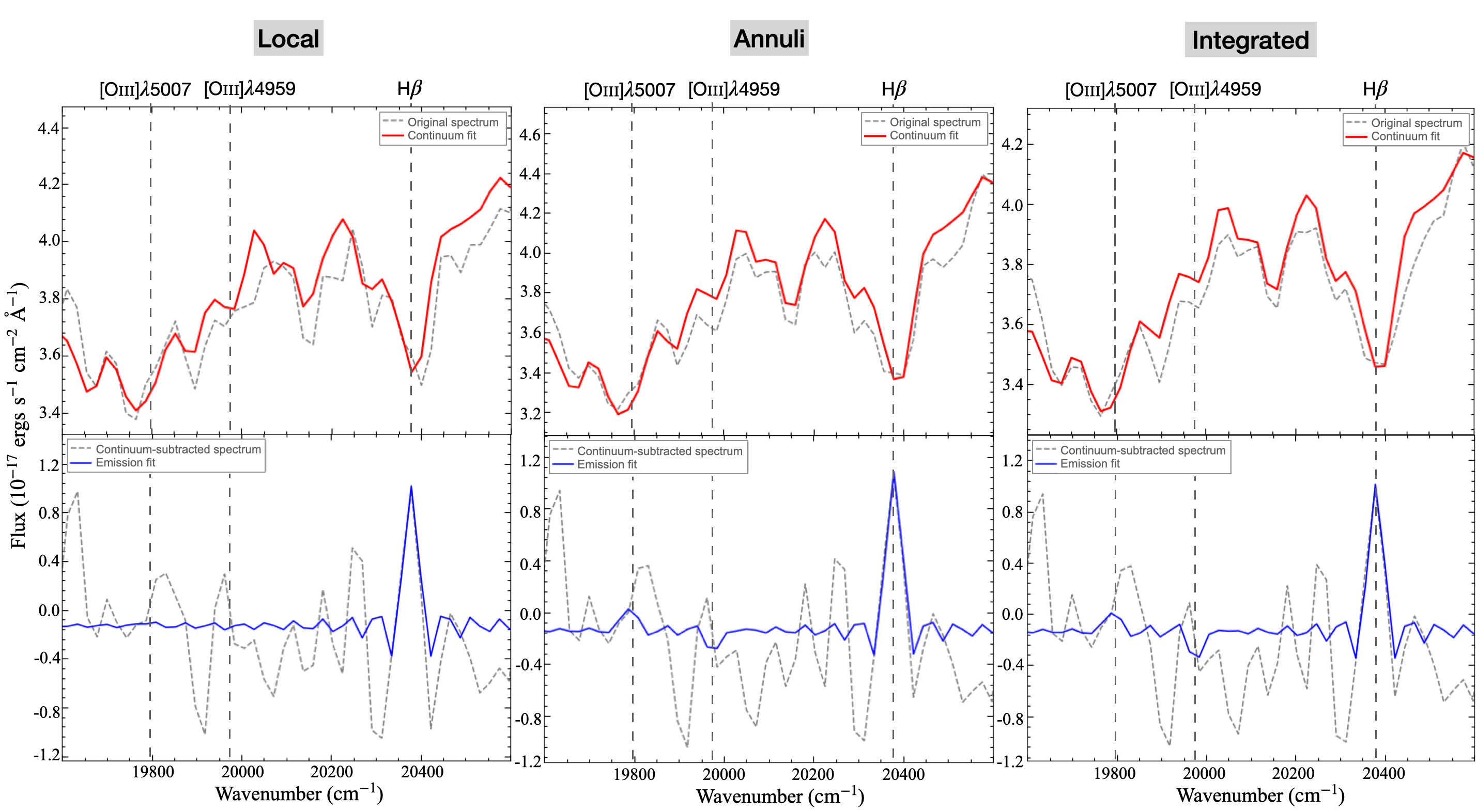}
  \caption{\label{fig:app_2} Demonstration of stellar continuum subtraction using three different integration methods, for the SN3 ({\it top}) and SN2 ({\it bottom}) data cubes. All spectra are extracted from a spaxel located at ${\rm RA~(J2000)}=01^{\rm h}24^{\rm m}47\fs8$ and ${\rm Dec.~(J2000)}=9^\circ32\arcmin23\farcs2$, $\approx600$~pc from the galaxy centre. {\it Left:} integration within $10\times10$ spaxels centred on the selected spaxel. {\it Middle:} integration within an annulus of width $150$~pc centred on the galaxy centre and encompassing the selected spaxel. {\it Right:} integration within $100\times100$ spaxels centred on the galaxy centre. In each panel, the top half shows the integrated spectrum (dashed gray curve) and the pPXF fit (solid red curve); the bottom half the residual spectrum (data - model; dashed grey curve) and the emission-line fit (solid blue curve). The black dashed vertical lines indicate the expected positions of the brightest emission lines, labelled above each plot.
  }
\end{figure*}

To subtract the stellar continuum (and thus the stellar absorption) co-spatial with the emission lines, we use the pPXF algorithm \citep{Cappellari2022} with the Medium-resolution Isaac Newton Telescope Library of Empirical Spectra (MILES; \citealt{sanchez2006MNRAS.371..703S}), 
while the emission lines are modelled using the SITELLE `sinc' line spread function implemented into the pPXF routines (as described in Mass\'{e} et al.\ in prep.).   
As the $S/N$ in any single spaxel is generally too low to obtain a reliable fit to the stellar continuum, it is necessary to combine several spaxels (for the stellar continuum subtraction only, not the emission-line fits). To optimise this process, we experimented with combining spaxels using three different methods: 1) $10\times10$ spaxels centred on each spaxel, 2) annuli of $150$~pc width centred on the galaxy centre
and 3) $100\times100$ spaxels centred on the galaxy centre. In all cases, the spectrum of each spaxel is shifted to the galaxy rest frame using the H$\alpha$ velocity map (Figure~\ref{fig:data}) prior to being added and subsequently fitted with pPXF. Examples of stellar continuum models generated by pPXF for these spectra are shown in Figure~\ref{fig:app_2}. The pPXF best-fitting models are then subtracted from every spaxel after being shifted to that spaxel's mean velocity and scaled to the correct continuum level. We show the corresponding emission-line best-fitting results using {\tt ORCS} in Figure~\ref{fig:app_2}. All three methods perform reasonably well to capture and correct for the absorption features. To obtain the most reliable emission-line measurements, we select the method that yields the highest $S/N$ at each spaxel. All the analyses discussed in this work are based on emission-line maps obtained in this manner.

\begin{figure*}
  \centering
  \includegraphics[width=0.9\textwidth]{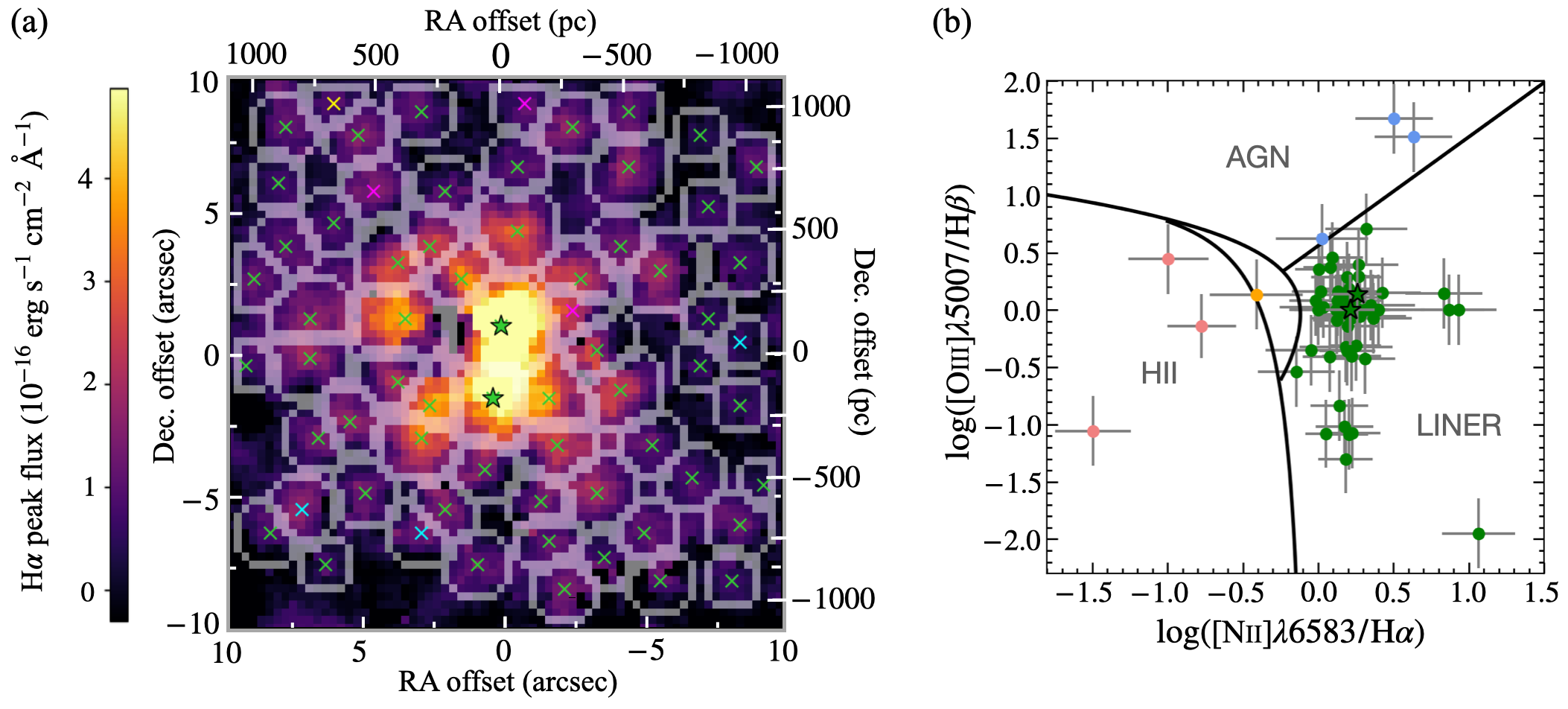}
  \includegraphics[width=0.9\textwidth]{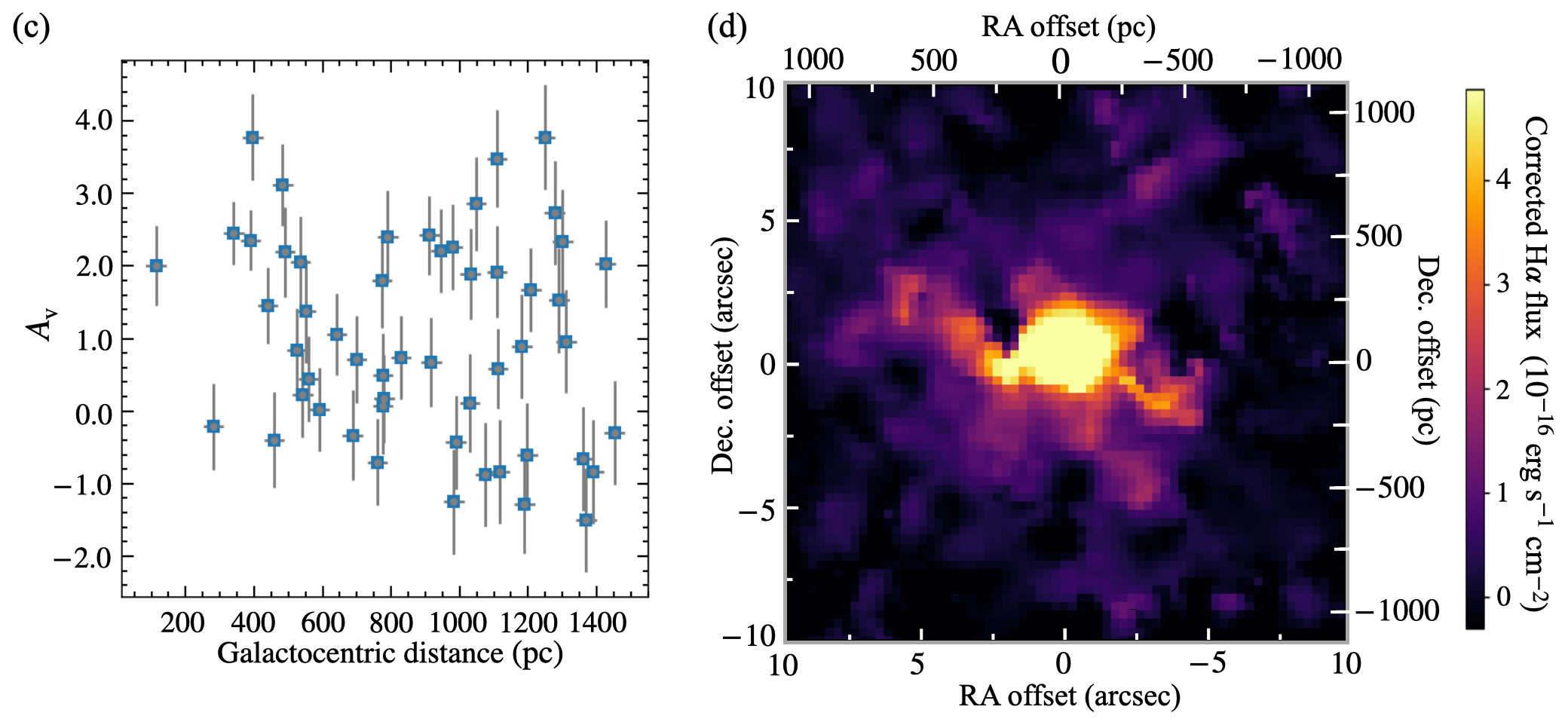}
  \caption{\label{fig:app_3} (a) NGC~524 H$\alpha$ peak flux map, overlaid with the identified emission-line regions (grey contours) and the locations of the identified emission local maxima (crosses; colour-coded according to the ionisation mechanism identified in panel~b). The two bright central regions are highlighted as stars. (b) Emission-line ratio diagnostic diagram of \citet{law2021sdss}, with the identified emission-line region data points overlaid (colour-coded according to the ionisation mechanism). Again, the two bright central regions of panel a) are highlighted as stars. 
  (c) Extinction ($A_{\rm v}$) of each emission-line region as a function of galactocentric distance. (d) Extinction-corrected H$\alpha$ surface brightness map. 
  }
\end{figure*}

To characterise the ionisation mechanisms of the ionised gas, emission-line regions must first be defined. To do this, the positions of the emission-line local maxima and their associated emission were identified using the algorithm described by Savard et al.\ (in prep.). To minimise the impact of noise, we use here the peak flux rather than the integrated flux of the H$\alpha$ emission line. The Laplacian of the H$\alpha$ peak flux map (corrected for absorption) was first used to identify local maxima in running $3\times3$ spaxels boxes, and only the maxima above a certain threshold were kept. To account for the H$\alpha$ background, this threshold is equal to $1.7$ times the H$\alpha$ peak flux noise calculated in larger $20\times20$ spaxels running boxes. The positions of these (remaining) maxima are shown on the H$\alpha$ peak flux map of Figure~\ref{fig:app_3}~(a). 

A `working zone' around each selected local maximum is then defined by assigning to each pixel a dominant maximum (the neighbouring maximum with the highest $a/r^2$ ratio, where $a$ is the peak flux of the maximum and $r$ the distance of the maximum from the pixel). The zone sizes vary from $\approx200$ to $\approx350$~pc (mean radius from the maximum to the zone's edge) depending on the maximum's peak flux. These zones are then adopted as our final emission-line regions, 
shown in Figure~\ref{fig:app_3}~(a). They are also transposed to the SN3 and SN2 data cubes, to calculate the integrated SN3 and SN2 spectra of all the regions and measure their emission lines with {\tt ORCS}. 

\begin{figure*}
   \centering
   \includegraphics[width=0.4\textwidth]{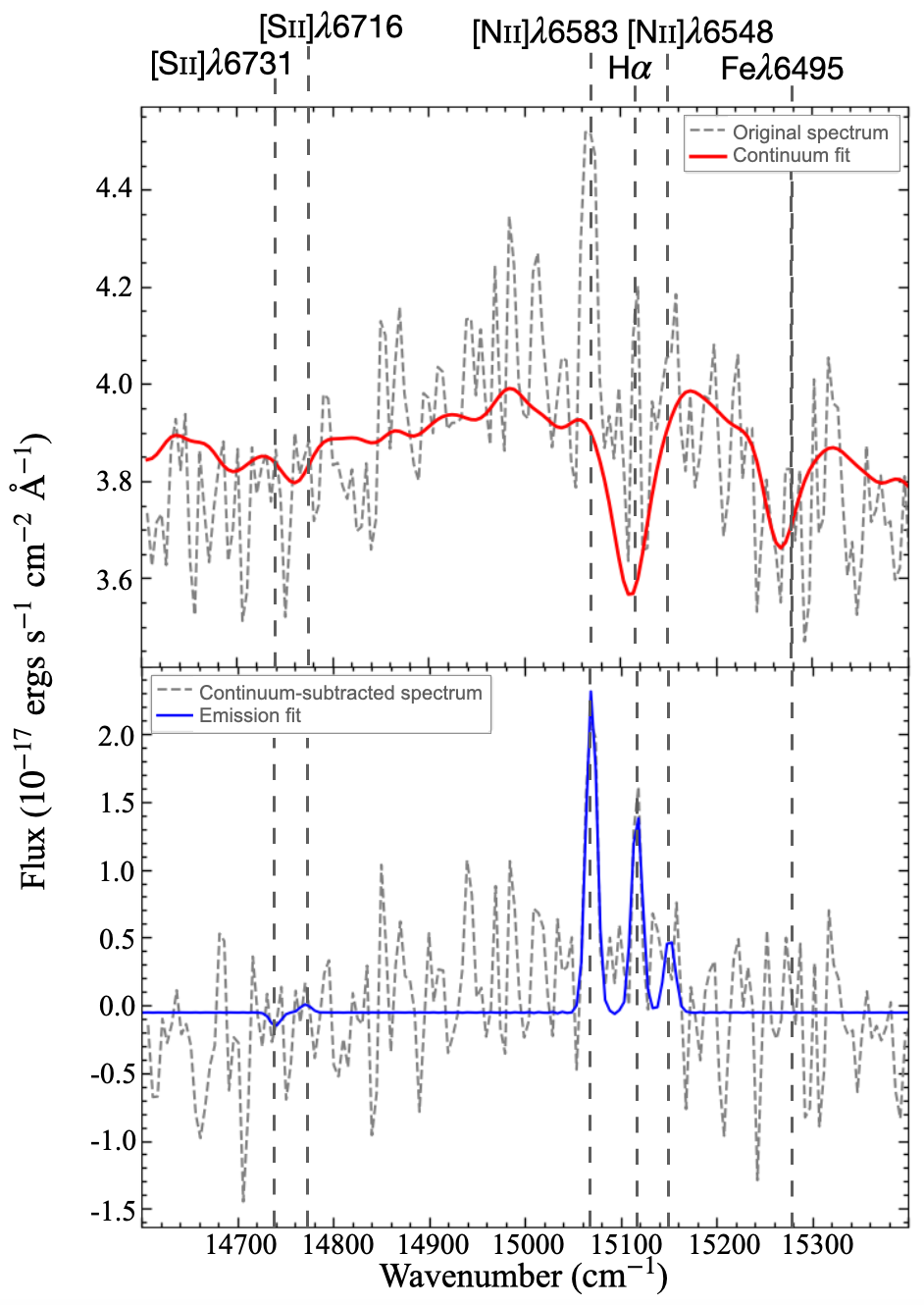}
   \caption{\label{fig:app_4} As the top-left panel of Figure~\ref{fig:app_2}, but for the spaxel at the centre of NGC~524. 
   }
\end{figure*}

We then attempt to distinguish different ionisation mechanisms by exploiting primarily emission-line ratios (so-called `BPT diagrams'; e.g.\ \citealt{1981PASP...93....5BPT}), using the most up-to-date diagnostics from \citet{law2021sdss} shown in Figure~\ref{fig:app_3}~(b). The H$\alpha$ emission local maxima shown in Figure~\ref{fig:app_3}~(a) are colour coded according to this diagnostic diagram. The emission of nearly all regions is characteristic of low-ionisation (nuclear) emission regions (LINERs). 
Although the central region of NGC~524 is very bright and has a compact radio source, we confirm from its emission spectrum (see Figure~\ref{fig:app_4}) that there is no significant AGN signature. The central spectrum is instead similar to that of the strong diffuse ionised-gas (DIG) component present away from the galaxy centre. Based on this, we infer that essentially all SFRs derived from H$\alpha$ emission are upper limits.

The extinction of each region is calculated considering the Balmer decrement, based on the integrated flux ratio of H$\alpha$ and H$\beta$. Figure~\ref{fig:app_3}~(c) shows the extinction ($A_{\rm v}$) calculated as a function of galactocentric distance. A few regions have negative $A_{\rm v}$, that can be explained by uncertainties of the flux calibration and/or the measurements, and/or the fact that the intrinsic H$\alpha$/H$\beta$ flux ratio can deviate from $2.87$ in diffuse ionised-gas regions. Nevertheless, all data points have $A_{\rm v}>0$ within the uncertainties. For each negative $A_{\rm v}$ region, we use the upper limit to correct for the extinction. The resulting extinction-corrected H$\alpha$ surface brightness map is shown in Figure~\ref{fig:app_3}~(d).   

\end{document}